\pdfoutput=1

\documentclass[11pt,twoside,a4paper,cmspaper,final,collab]{cms-tdr}
\def\svnVersion{493428P}\def\cmsCernNoTag{CERN-EP-2018-279}\def\cmsCernDate{\today}\def\cmsMessage{Published in the Journal of High Energy Physics as \href{http://dx.doi.org/10.1007/JHEP03(2019)127}{\doi{10.1007/JHEP03(2019)127.}}}

\begin{document}\cmsNoteHeader{B2G-18-001}

\hyphenation{had-ron-i-za-tion}
\hyphenation{cal-or-i-me-ter}
\hyphenation{de-vices}
\RCS$Revision: 488613 $
\RCS$HeadURL: svn+ssh://svn.cern.ch/reps/tdr2/papers/B2G-18-001/trunk/B2G-18-001.tex $
\RCS$Id: B2G-18-001.tex 488613 2019-02-08 19:11:34Z knash $

\newlength\cmsFigWidth
\ifthenelse{\boolean{cms@external}}{\setlength\cmsFigWidth{0.85\columnwidth}}{\setlength\cmsFigWidth{0.4\textwidth}}
\ifthenelse{\boolean{cms@external}}{\providecommand{\cmsLeft}{upper\xspace}}{\providecommand{\cmsLeft}{left\xspace}}
\ifthenelse{\boolean{cms@external}}{\providecommand{\cmsRight}{lower\xspace}}{\providecommand{\cmsRight}{right\xspace}}
\newlength\cmsTabSpace\setlength\cmsTabSpace{1ex}

\newcommand{\TF}{\ensuremath{F(\pt,\eta)}\xspace}
\newcommand{\TFV}{\ensuremath{F_{\mathrm v}(\pt,\eta)}\xspace}
\newcommand{\tpr}{\ensuremath{\mathrm{T}}\xspace}
\newcommand{\bpr}{\ensuremath{\mathrm{B}}\xspace}
\newcommand{\mthb}{\ensuremath{m_{\mathrm{\cPqt\PH\cPqb}}}\xspace}
\newcommand{\cott}{\ensuremath{ \cot(\theta_2 )}}
\newcommand{\cottsq}{\ensuremath{ \mathrm{cot^2}(\theta_2 )}}
\providecommand{\CL}{CL\xspace}

\cmsNoteHeader{B2G-18-001}
\title{Search for a \PWpr boson decaying to a vector-like quark and a top or bottom quark in the all-jets final state}

\date{\today}

\abstract{
A search for a  heavy $\PWpr$ resonance decaying to one $\bpr$ or $\tpr$ vector-like quark and a top or bottom quark,
respectively, is presented.  The search uses proton-proton collision data collected in 2016 with the CMS detector at the LHC, corresponding to an integrated luminosity of 35.9\fbinv at $\sqrt{s}=13\TeV$. Both decay channels result in a final state with a top quark, a Higgs boson, and a $\cPqb$ quark, each produced with significant energy.
The all-hadronic decays of both the Higgs boson and the top quark are considered.
The final-state jets, some of which correspond to merged decay products of a boosted top quark and a Higgs boson,
are selected using jet substructure techniques, which help to suppress standard model backgrounds.
A $\PWpr$ boson signal would appear as a narrow peak in the invariant mass distribution of these jets.
No significant deviation in data with respect to the standard model background predictions is observed.
Cross section upper limits on $\PWpr$ boson production in the top quark, Higgs boson, and $\cPqb$ quark decay mode
are set as a function of the $\PWpr$ mass, for several vector-like quark mass hypotheses. These are the first limits for $\PWpr$ boson production in this decay channel, and cover a range of 0.01 to 0.43$\unit{pb}$ in the $\PWpr$ mass range between 1.5 and 4.0\TeV.}

\hypersetup{
pdfauthor={CMS Collaboration},
pdftitle={Search for a W' boson decaying to a vector-like quark and a top or bottom quark in the all-jets final state},
pdfsubject={CMS},
pdfkeywords={CMS, physics, B2G, resonances, Wprime}}

\maketitle

\section{Introduction}
\label{sec:introduction}

Many extensions of the standard model (SM) predict new massive charged gauge bosons
{\cite{doi:10.1146/annurev.nucl.55.090704.151502,PhysRevD.64.035002,PhysRevD.11.566}}.
The $\PWpr$ boson is a hypothetical heavy partner of the SM $\PW$ gauge boson that could be produced in
proton-proton (pp) collisions at the CERN LHC.  Searches for $\PWpr$ bosons have been most recently performed at a center-of-mass energy of $13\TeV$ by
the CMS and ATLAS Collaborations in the lepton-neutrino~{\cite{wpCMS1,Aaboud:2017efa}}, diboson~{\cite{Sirunyan:2018iff,wpAtlas1}},
and diquark~{\cite{wpCMS3,wpAtlas2}} final states.  Vector-like quarks (VLQs) are
hypothetical heavy partners of SM quarks for which the left- and right-handed
chiralities transform the same way under SM gauge groups.  Searches for VLQs have been performed by the CMS and ATLAS
Collaborations in both the single~\cite{vlqs1,vlqs2,vlqs3,vlqs4} and pair production~\cite{vlqp1,vlqp2,vlqp3} channels.
  The decay of the $\PWpr$ boson to a heavy $\bpr$ or $\tpr$ VLQ and a top or $\cPqb$ quark,
respectively, is predicted, \eg, in composite Higgs boson models with custodial symmetry protection~\cite{AGASHE2005165,Barducci2013,Barducci2016}.
These models stabilize the quantum corrections to the Higgs mass and preserve naturalness.  The $\PWpr$ branching fraction to a quark and a VLQ depends
on the VLQ mass, with a maximum of 50\% in the high VLQ mass range at the threshold of custodian production (see Ref.~\cite{Vignaroli:2014bpa}).

\begin{figure}[h]
\centering
\includegraphics[width=0.35\textwidth]{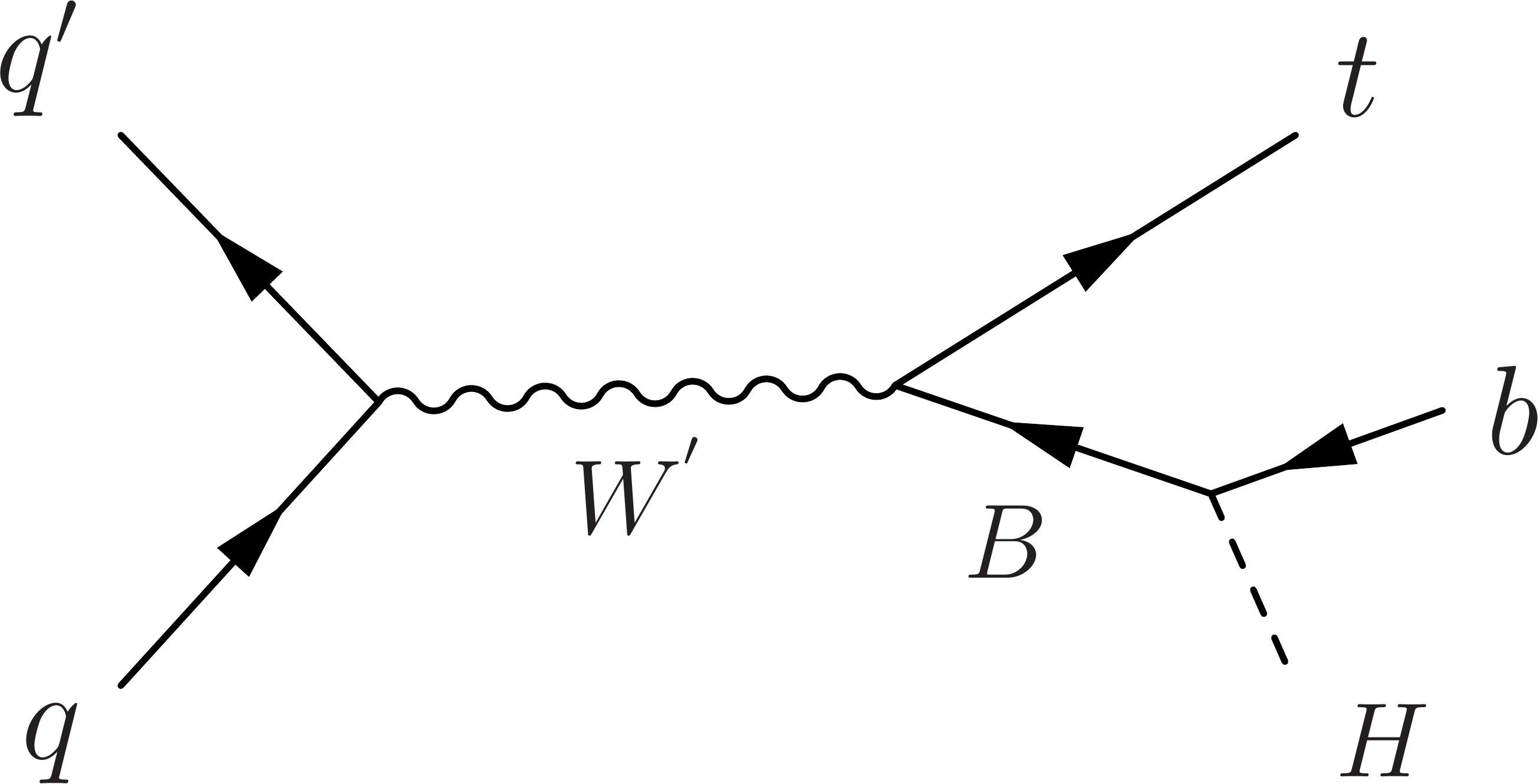}
\hspace{0.15\textwidth}
\includegraphics[width=0.35\textwidth]{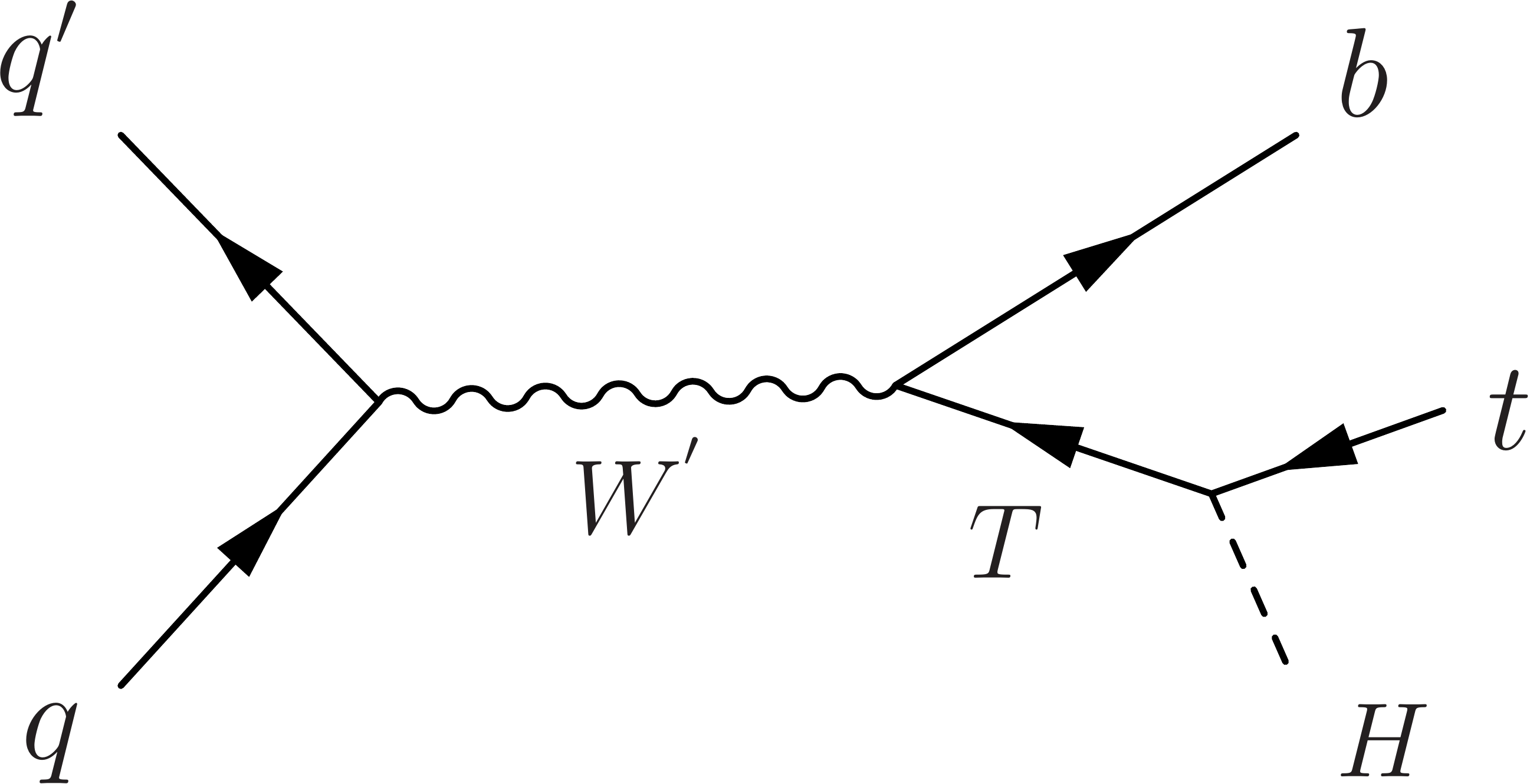}\\
\caption{
The $\PWpr$ boson production and decays considered in the analysis.
The analysis assumes equal branching fractions for $\PWpr$ boson to $\cPqt\bpr$ and $\cPqb\tpr$ and 50\% for each VLQ to q\PH.}
\label{figs:feyn}

\end{figure}

A search for a $\PWpr$ boson in this decay mode is presented for the first time.  The analysis
considers the decay channel where the $\bpr$ or $\tpr$ VLQ decays into a Higgs boson and a $\cPqb$ or top quark, respectively, in the
all-jets final state.
Both the $\bpr$ and $\tpr$ VLQ-mediated decays result in the same signature, as can be seen in Fig.\ref{figs:feyn}.
Because of the high $\PWpr$ and VLQ masses considered in this analysis, the decay products are highly Lorentz boosted.
These boosted decay products are reconstructed as single jets with distinct substructure,
which is used in the analysis to distinguish them from SM multijet production.
An inclusive search for a $\PWpr$ boson decaying to a top quark, a Higgs boson, and a $\cPqb$ quark is performed.
The SM background is dominated by events comprised
of jets produced via the strong interaction, referred to as quantum chromodynamics (QCD)
multijet events, and top quark pair production ($\ttbar$) events.  These backgrounds are
modeled by a combination of Monte Carlo (MC) simulation and control regions in data.
The invariant mass distribution of the three-jet system, $\mthb$, is used to set
the first limits on the $\PWpr$ boson production cross section in the decay channel to a $\bpr$ or $\tpr$ VLQ.
The data sample used in the analysis corresponds to an integrated luminosity of 35.9$\fbinv$~\cite{CMS-PAS-LUM-17-001}
of pp collision data at $\sqrt{s}=13\TeV$, recorded in 2016.

The theoretical framework followed in the analysis is described in Ref.~\cite{Vignaroli:2014bpa}.
In this model the top and $\PWpr$ are superpositions of elementary and composite modes, with
the top degree of compositeness given by $s_{\mathrm{L}}$, and the mixing angle of the elementary and composite $\PWpr$ states
given by $\theta_{2}$.
The $\PWpr$ boson production cross section is inversely proportional to $\cottsq$,
 but low $\cott$ values tend to be dominated by the leptonic $\PWpr$ boson decay mode.  High values of the $s_{\mathrm{L}}$ parameter increase the
relative phase space for the decay into two VLQs, whereas low $s_{\mathrm{L}}$ values enhance the $\PWpr$ diboson decays.
The analysis assumes this theoretical framework as evaluated at $s_{\mathrm{L}}=0.5$ and $\cott=3$, which is chosen for the purposes of
sensitivity in the $\PWpr$ decay channel to a single VLQ.
The expected signal cross sections in the analysis are evaluated at 13\TeV using the framework of Ref.~\cite{Vignaroli:2014bpa} for $\PWpr$ masses in the range
1.5 to 4.0\TeV with the
assumptions that the $\PWpr\to$VLQ branching fraction is equally distributed
between the t$\bpr$ and b$\tpr$ final states.  As a benchmark for the analysis, the VLQ branching fractions
for each of the decays  $\bpr\to$\cPqb\PH and $\tpr\to$\cPqt\PH are assumed to be 50\%, consistent with the benchmark used in other recent searches.

\section{The CMS detector}
The central feature of the CMS apparatus is a superconducting solenoid of 6\unit{m} internal diameter,
providing a magnetic field of 3.8\unit{T}. Within the solenoid volume are a silicon pixel and strip tracker,
a lead tungstate crystal electromagnetic calorimeter (ECAL), and a brass and scintillator hadron calorimeter (HCAL),
each composed of a barrel and two endcap sections. Forward calorimeters extend the pseudorapidity coverage provided by the
barrel and endcap detectors. Muons are detected in gas-ionization chambers embedded in the steel flux-return yoke outside the solenoid.
A more detailed description of the CMS detector, together with a definition of the coordinate system used and the
relevant kinematic variables, can be found in Ref.~\cite{Chatrchyan:2008zzk}.

The particle-flow algorithm~\cite{CMS-PRF-14-001} aims to reconstruct and identify each individual particle with an optimized
combination of information
from the various elements of the CMS detector. The energy of each photon is obtained from the ECAL measurement. The energy of each electron
is determined from a combination of the electron momentum at the
primary interaction vertex as determined by the tracker, the energy of the corresponding ECAL cluster, and the energy sum of all
bremsstrahlung photons spatially compatible with originating from the electron track. The energy of each muon is obtained from the momentum,
which is measured by the curvature of the corresponding track. The energy of each charged hadron is determined from a combination of their momentum measured
in the tracker and the matching ECAL and HCAL energy deposits, corrected for zero-suppression effects and for the response function
of the calorimeters to hadronic showers. Finally, the energy of each neutral hadron is obtained from the corresponding corrected
ECAL and HCAL energies that are not associated with a charged hadron track.

Jets are clustered with the
anti-$\kt$ \cite{Cacciari:2008gp} algorithm in the $\FASTJET$ 3.0 \cite{Cacciari:2011ma} software package.
Jet momentum is determined as the vectorial sum of all particle momenta in the jet, and is found from simulation to be within
5 to 10\% of the true momentum over the whole \pt spectrum and detector acceptance. Additional pp interactions within
the same or nearby bunch crossings (pileup) can contribute additional tracks and calorimetric energy depositions to the jet momentum.
To mitigate this effect, charged particles originating from sub-leading pp collision vertices within the same or adjacent
bunch crossings are discarded in the jet clustering procedure, where the primary collision vertex is defined as the
vertex largest quadrature-summed $\pt$ of all reconstructed particles.
To account for the neutral pileup component, the pileup per particle identification (PUPPI)
algorithm \cite{Bertolini2014} is used, which applies weights that rescale the jet transverse momentum based on the per-particle
probability of originating from the primary vertex prior to jet clustering.
Jet energy corrections are derived from simulation studies so that the average measured response of jets becomes identical
to that of particle level jets. In situ measurements of the momentum balance in dijet, photon+jet, Z+jet, and multijet events are
used to determine any residual differences between the jet energy scale in data and in simulation, and appropriate corrections are made~\cite{Khachatryan:2016kdb}.
Additional selection criteria are applied to each jet to remove jets potentially dominated by instrumental effects or reconstruction failures.
The jet energy resolution amounts typically to 15\% at 10\GeV, 8\% at 100\GeV, and 4\% at 1\TeV, to be compared to
about 40, 12, and 5\% obtained when the calorimeters alone are used for jet clustering.

Events of interest are selected using a two-tiered trigger system~\cite{Khachatryan:2016bia}.
The first level (L1), composed of custom hardware processors, uses information from the calorimeters and muon detectors to
select events at a rate of around 100\unit{kHz} within a time interval of less than 4\mus. The second level,
known as the high-level trigger (HLT), consists of a farm of processors running a version of the full event reconstruction
software optimized for fast processing, and reduces the event rate to around 1\unit{kHz} before data storage.

\section{Simulated samples}
\label{sec:datasample}
The $\ttbar$ production background is estimated from simulation, and is generated with
\POWHEG2.0~\cite{Powheg,Alioli:2010xd,Nason:2004rx,Frixione:2007nw}.  The signal samples
are generated at leading order using \MGvATNLO 2.3.3~\cite{Alwall:2014hca,Alwall:2007fs}
with the NNPDF3.0 leading order parton distribution function (PDF) set, in the mass range from 1.5 to 4.0\TeV in 0.5\TeV increments.
The analysis uses a QCD multijet sample as a cross check for the background estimate, which is also generated at LO with \MGvATNLO.
Parton showering and hadronization are simulated with $\PYTHIA$8.212~\cite{Sjostrand:2014zea} using either the CUETP8M2T4 \cite{CMS-PAS-TOP-16-021} or CUETP8M1 \cite{Khachatryan:2015pea} underlying event tunes.
For each $\PWpr$ boson mass point, three VLQ mass points are generated with the VLQ mass range from 0.8 to 3.0\TeV.
The generated VLQ masses are scaled to the $\PWpr$ boson mass ($m_{\PWpr}$) such that
there is a low ($\approx$1/2 $m_{\PWpr}$),  medium ($\approx$2/3 $m_{\PWpr}$), and high ($\approx$3/4 $m_{\PWpr}$) mass
sample for each $\PWpr$ boson mass point in order to explore the sensitive phase space of the boosted $\PWpr$ boson decay products.
The generated $\PWpr$ boson and VLQ widths are narrow as compared with the detector and reconstruction resolutions which is in accord with
theoretical predictions for most of the analyzed phase space.
The simulation of the CMS detector uses $\GEANTfour$~\cite{g4c}.
All MC samples include pileup simulation and are weighted such that the distribution of the number of interactions per bunch
crossing agrees with that observed in data.

\section{Event reconstruction}
\label{sec:eventreconstruction}

The $\PWpr\to\tpr/\bpr\to\cPqt\PH\cPqb$ channel is characterized by three high-$\pt$ jets.
The jets from the top quark (top jet) and Higgs boson (Higgs jet) decays tend to be wide and massive, whereas the
jet from the $\cPqb$ quark ($\cPqb$ jet) will tend to be narrow and have a lower mass.   Therefore, one jet clustered with the
anti-$\kt$ algorithm with a distance parameter of 0.8 (AK8 jet)
with $\pt>300\GeV$ is required for the Higgs boson candidate jet.  One AK8 jet with $\pt>400\GeV$ is required for the top quark candidate jet.
One anti-\kt jet with a distance parameter of 0.4 (AK4 jet) with $\pt>200\GeV$ is required for the $\cPqb$ candidate jet.
The separation $\Delta$R ($\sqrt{\smash[b]{(\Delta\phi)^2+(\Delta\eta)^2}}$) between the two AK8 jets is required to be at least 1.8 in order to
reduce the correlation of jet shapes arising from the abutting of jet boundaries, which can bias the background estimate.
The AK8 jets are then selected as being consistent with a top quark or a Higgs boson decay using the tagging procedures defined below.
The collection of jets considered for the $\cPqb$ quark candidate is then populated by AK4 jets with $\Delta$R of at least 1.2 from the tagged AK8 jets. In the case of
multiple jets with the same tag, the tagged candidate is chosen randomly.
Jet identification criteria are used for these three jets in order to reduce the impact of
spurious jets from detector noise~\cite{CMS-PAS-JME-16-003}.
All jets in the analysis are required to be within $\abs{\eta}<2.4$.

\label{sec:Trigger}

Because the signal of interest is a high mass resonance decaying to multiple high-$\pt$ jets, data events are triggered
by $\HT>$800 or 900$\GeV$, where $\HT$ is defined as the sum of all AK4 jet $\pt$ in the event, or a AK8 jet $\pt>450\GeV$.  The signal of interest usually
fulfills the high $\HT$ trigger requirement, 
and the AK8 jet $\pt$ trigger is included to overcome 
an issue in the trigger $\HT$ calculation that impacts about 24\% of the analyzed data.

\begin{figure}[htb]
\centering
\includegraphics[width=0.75\textwidth]{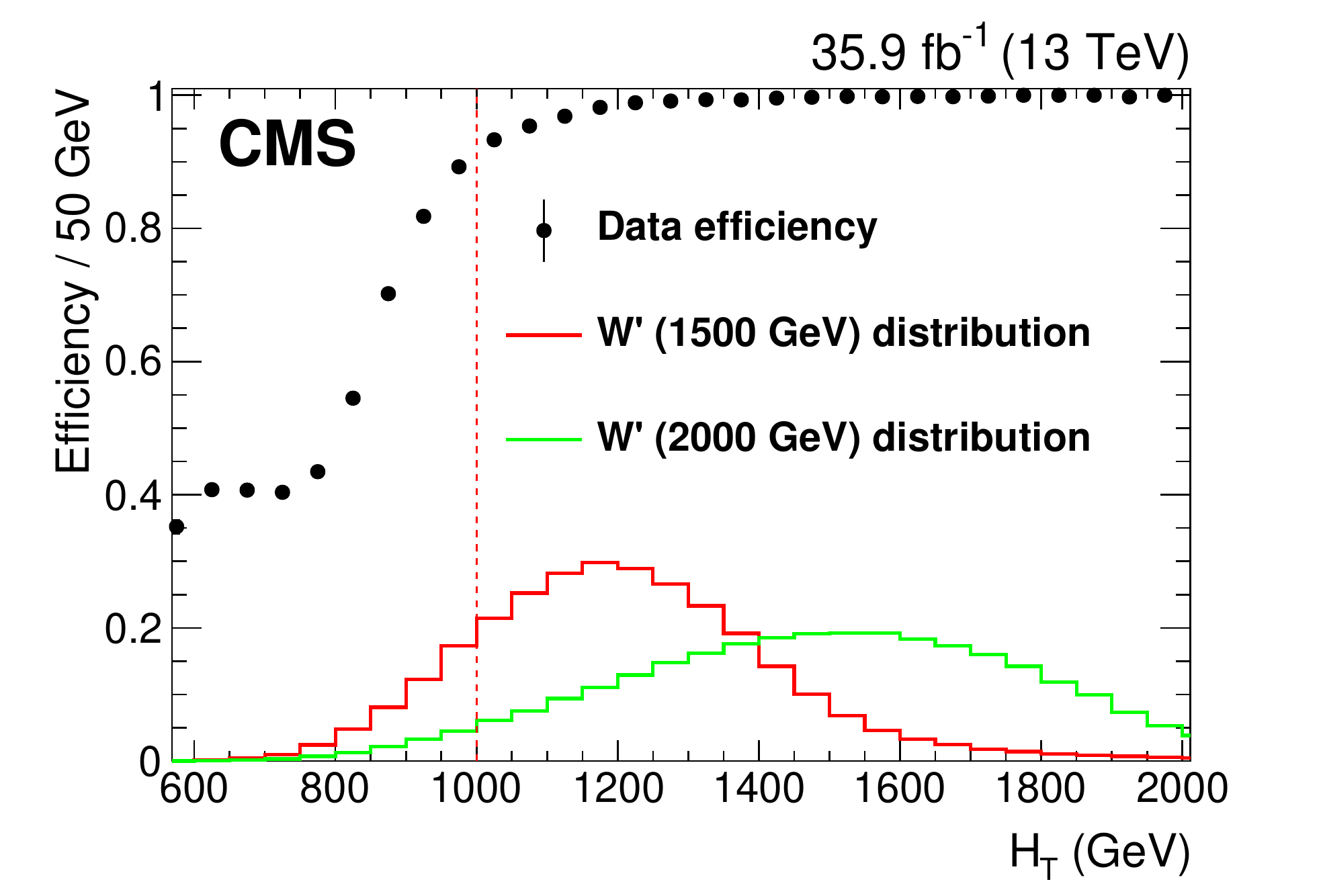}
\caption{
Trigger efficiency as a function of $\HT$.  Events are required to have $\HT>1\TeV$
as is indicated by the red dashed line.  The $\HT$ distributions of two $\PWpr$ signal hypotheses are shown for comparison, normalized to
unit area.
}
\label{figs:Trigger}
\end{figure}

The efficiency of the trigger selection is studied using a sample of events that have at least one muon of $\pt > 24\GeV$.
The fraction of these events that pass the full trigger selection is defined as the trigger efficiency and is shown in Fig.~\ref{figs:Trigger}
as a function of $\HT$. The offline event selection requires that $\HT$ be larger than 1\TeV which ensures that the trigger efficiency is
larger than 93\% near the threshold and is nearly 100\% over most of the signal region.
Although there is little inefficiency due to the trigger, this is taken into account as an event weight when
processing MC samples.

\subsection{Top jet tagging}
\label{sec:toptag}

For top quarks with $\pt>400\GeV$, the decay products, one $\cPqb$ quark and two light quarks,
can merge into a single AK8 jet.  Top quark jets are identified using a set of three quantities defined below.

The N-subjettiness~\cite{Thaler:2011gf} algorithm defines the $\tau_{N}$ variable, which quantifies how consistent the jet energy pattern is with
$N$ or fewer hard partons, with the low $\tau_{N}$ values being more consistent with $N$ or fewer partons.  In the case of a
top quark hadronic decay, the ratio of $\tau_3$ to $\tau_2$ is used.

The merged top jet can also be discriminated from background by using the large top quark mass. The modified mass drop tagger algorithm~\cite{Dasgupta2013},
also known as the ``soft drop'' algorithm \cite{Larkoski:2014wba} with $\beta=0$
and $z=0.1$ is used to calculate this mass variable, $m_{\mathrm{SD}}^\cPqt$.
This algorithm declusters the jet, and removes soft radiations, thus allowing a
clearer separation between the merged top jet and background.

Finally, as the top jet contains a $\cPqb$ quark, additional discrimination power can be achieved by using subjet $\cPqb$ tagging
with the combined secondary vertex version 2 (CSVv2) $\cPqb$ tagging algorithm ($SJ_{\mathrm{csvmax}}$)~\cite{Sirunyan:2017ezt}.  We use a $\cPqb$ tagging operating point
defined by a 10\% misidentification probability with approximately an 80\% efficiency.

The MC to data correction (scale factor) for the top tagging operating point in Table~\ref{table:regions} is measured to be $1.07^{+0.11}_{-0.04}$ in a sample enriched
in semileptonic $\ttbar$ production, using the same procedure as outlined in Ref.~\cite{CMS-PAS-JME-16-003}.

\subsection{Higgs jet tagging}
\label{sec:htag}
In the case of a highly boosted Higgs boson in the $\bbbar$ decay mode, the decay products tend to merge into a jet that has a mass
consistent with a Higgs boson and that contains two $\cPqb$ hadrons clustered into the jet.  Once again, the
soft drop algorithm is used to provide the variable $m_{\mathrm{SD}}^\PH$ as a measure of the Higgs boson jet mass, but in this case the jet is scaled using a simulation-derived correction suitable for
resonances below the top jet tagging mass window~\cite{PhysRevD.97.072006}, which is $\pt$ and $\eta$ dependent but results in
a 5-10\% mass amplification in both data and MC.
Scale factors are used for the jet mass scale and resolution, which are derived from a fit to the distribution of the $\PW$ boson
jet $m_{\mathrm{SD}}^\PH$ spectrum in
a sample enriched in semileptonic $\ttbar$ production using the technique outlined in Ref.~\cite{CMS-PAS-JME-16-003}.

To identify the two $\cPqb$ quarks clustered into the merged Higgs jet, a dedicated double-$\cPqb$ tagging algorithm (Dbtag) is used at an operating point
with a misidentification probability of approximately 3\% and an efficiency of 50\%.
Data samples enriched in QCD produced $\bbbar$ and $\ttbar$ events are used to establish scale factors for this tagger for the cases of signal
and mistagged top quarks, respectively~\cite{Sirunyan:2017ezt}.

\begin{figure}[htp]
\centering
\includegraphics[width=0.48\textwidth]{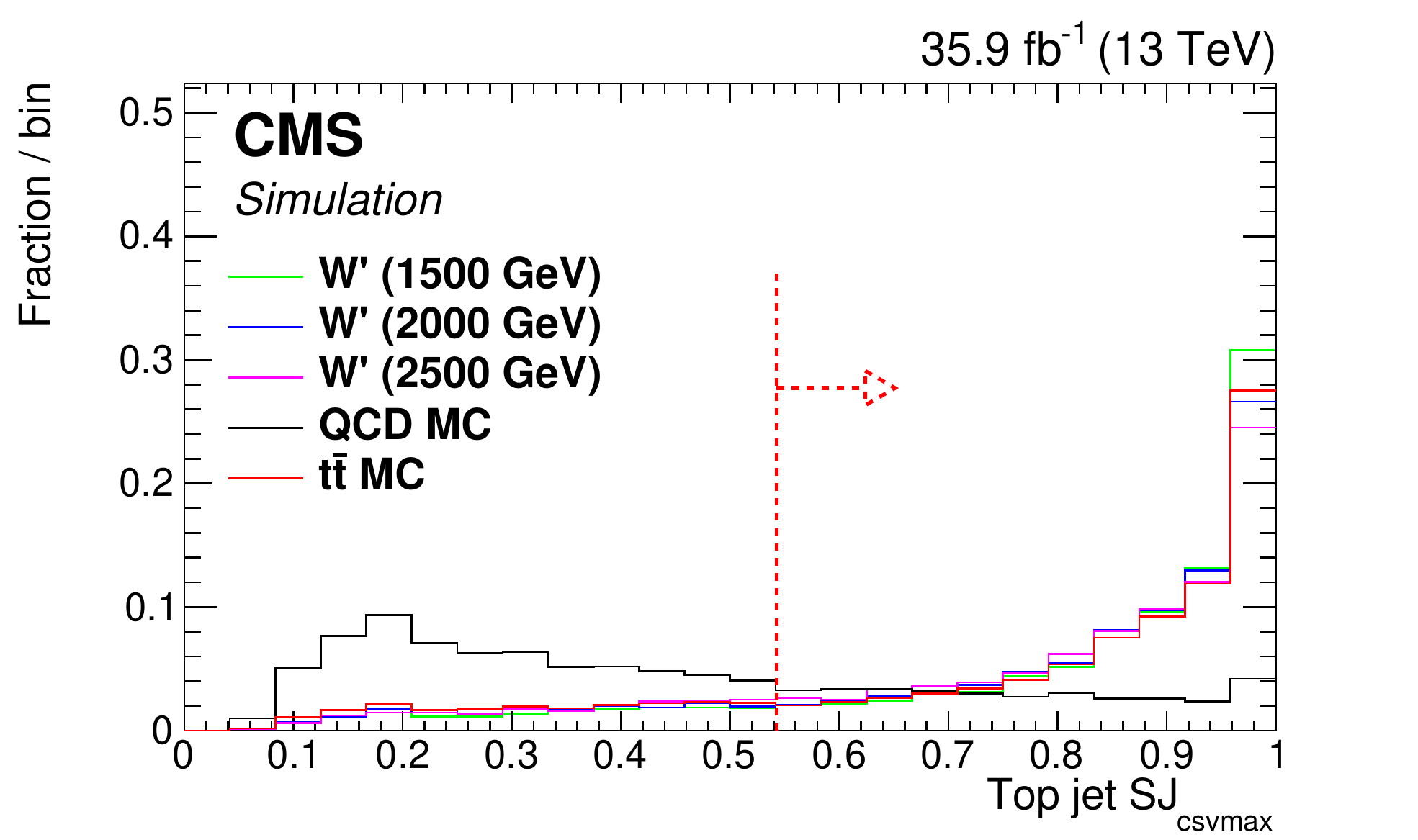}
\includegraphics[width=0.48\textwidth]{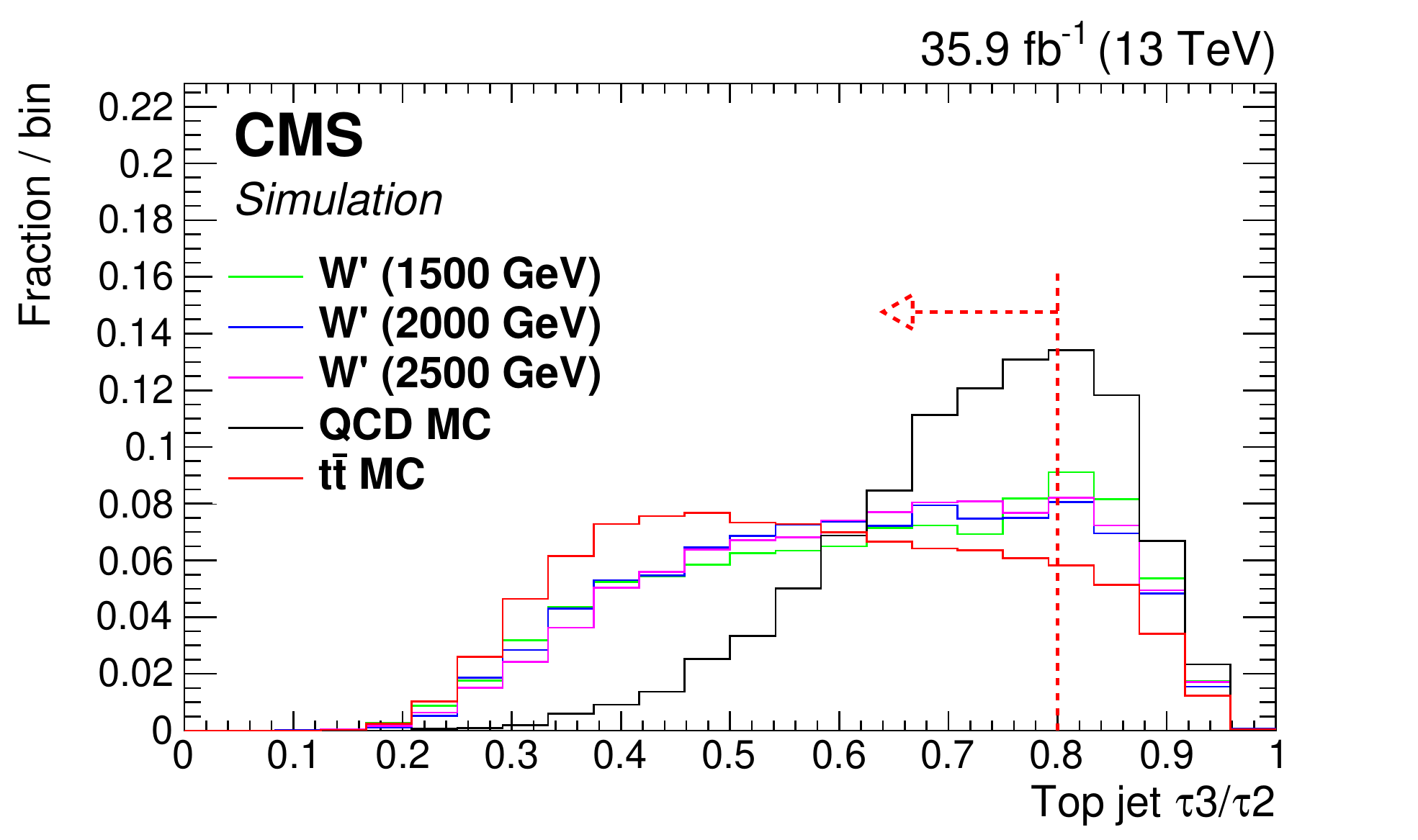}\\
\includegraphics[width=0.48\textwidth]{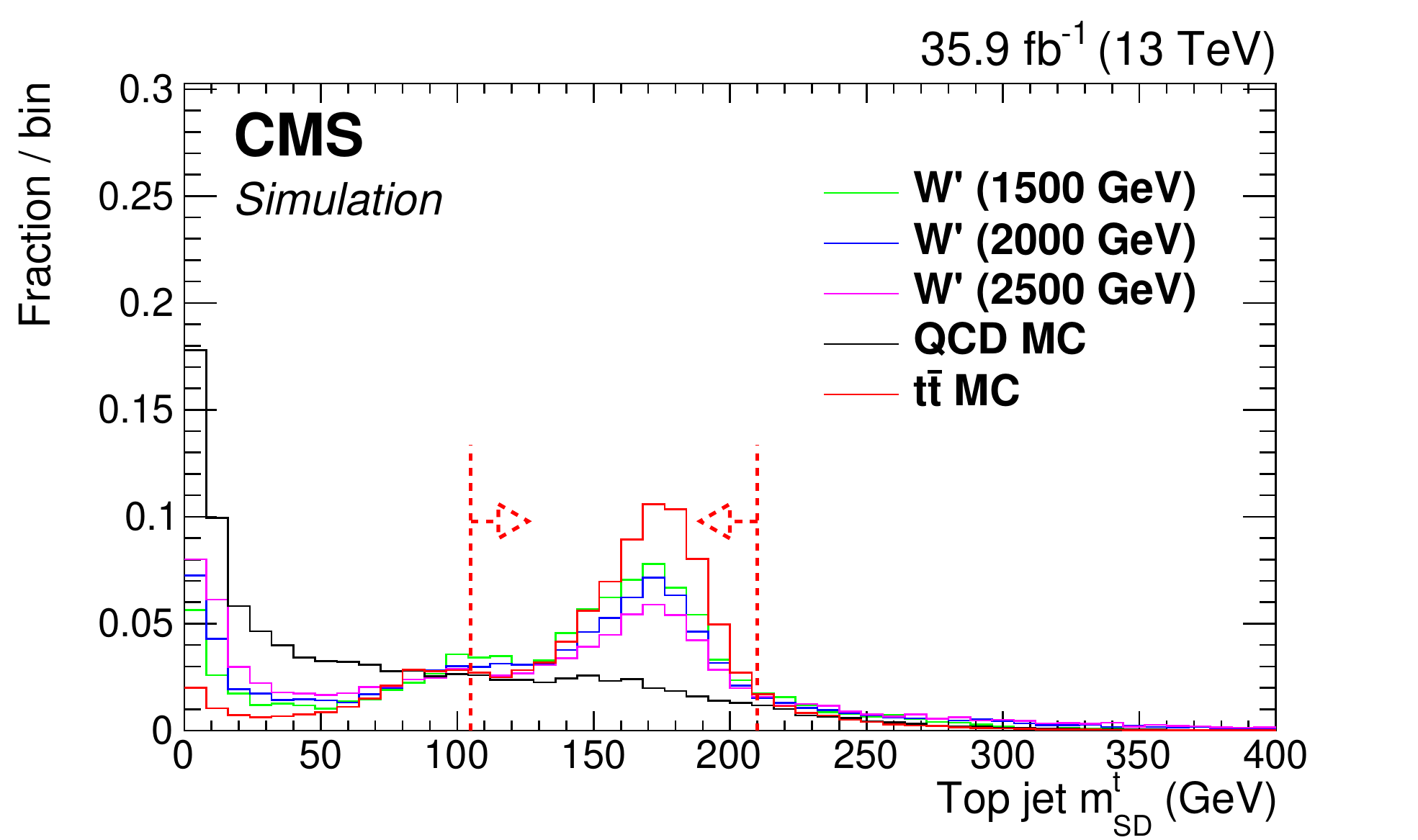}
\includegraphics[width=0.48\textwidth]{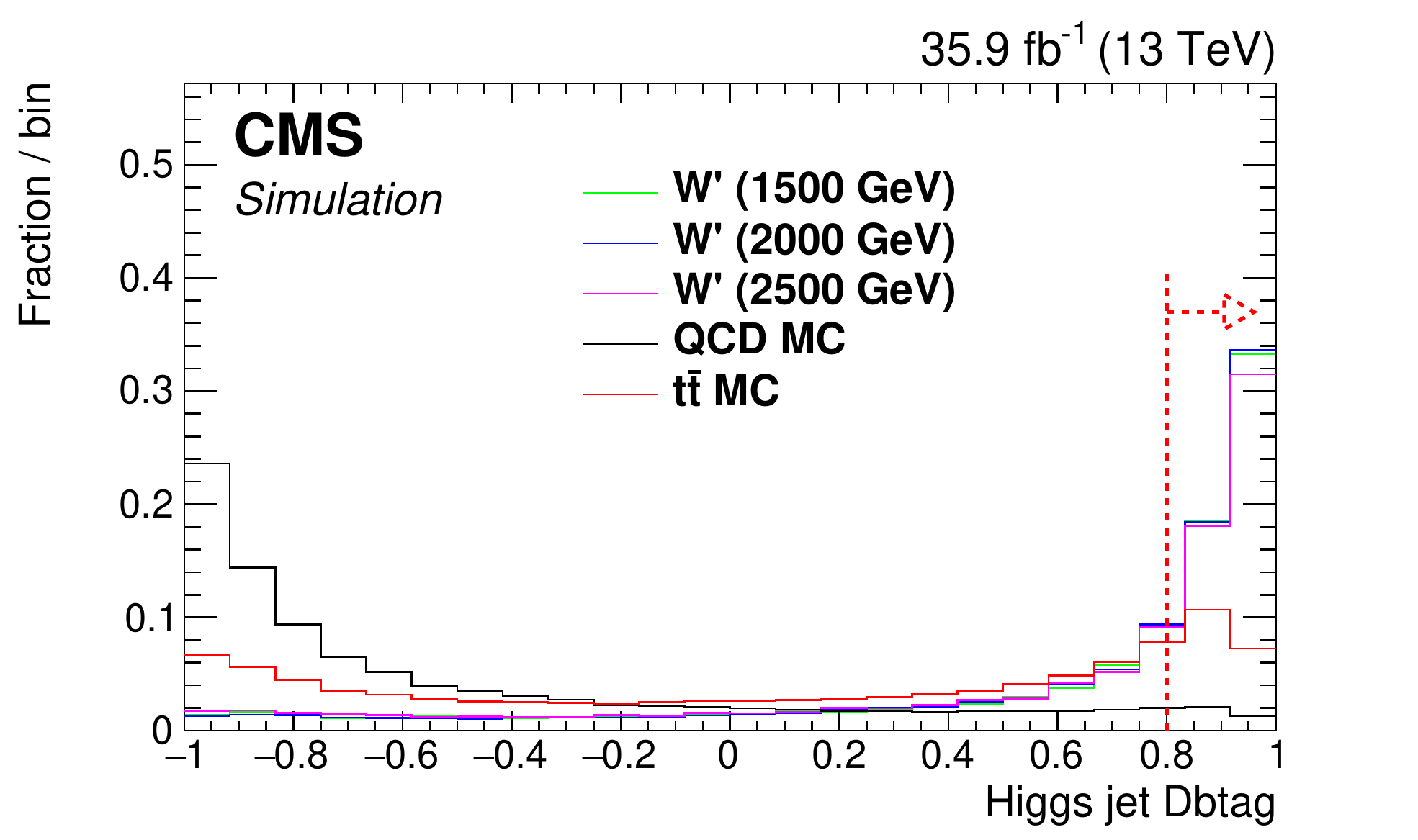}\\
\includegraphics[width=0.48\textwidth]{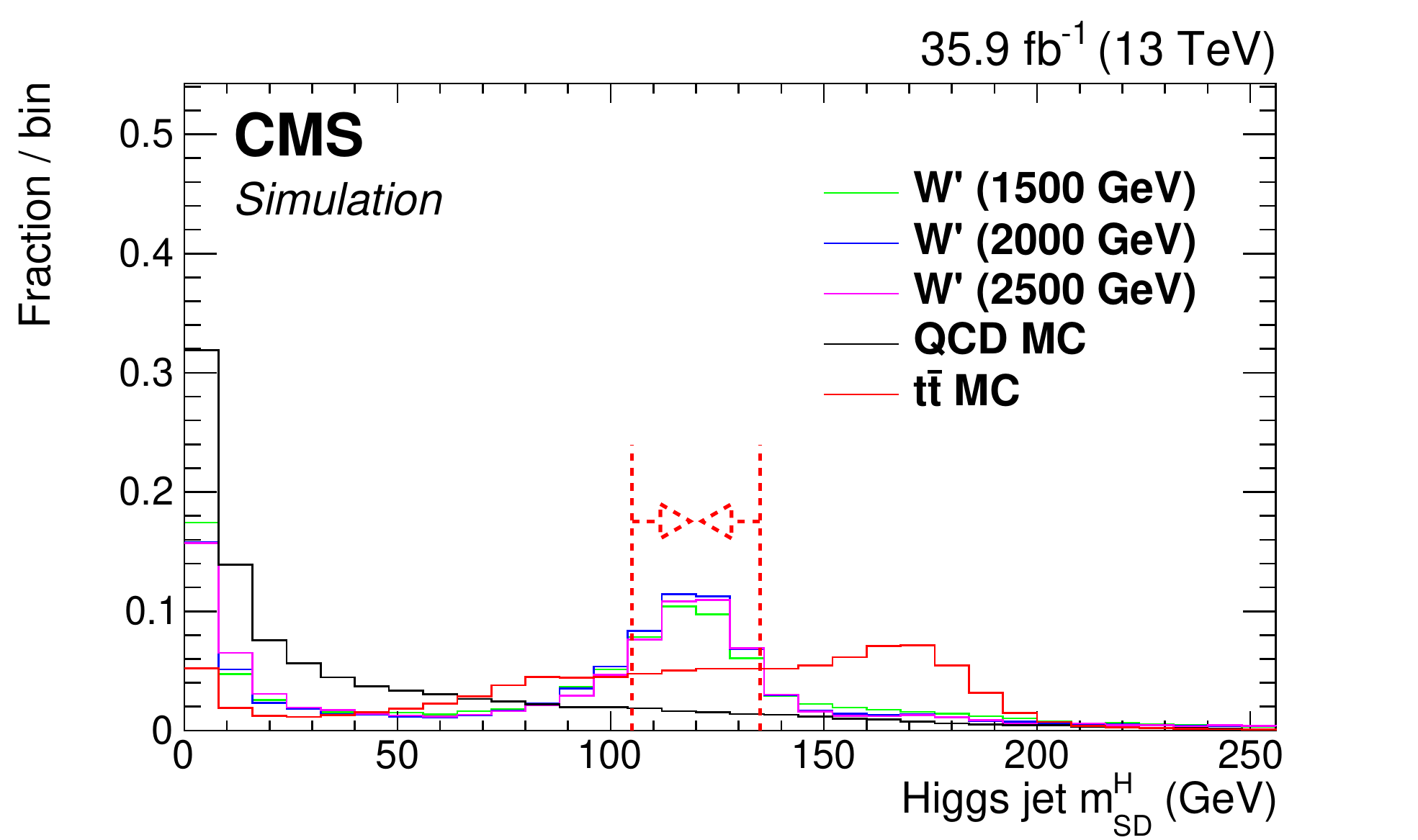}
\caption{
Normalized distributions of the discriminating variables in $\ttbar$, QCD, and signal MC simulation.
The distributions shown, from upper left to lower right,
are of the variables: the maximum subjet $\cPqb$ tag, $\tau_3$/$\tau_2$, and $m_{\mathrm{SD}}^\cPqt$, all used for top quark discrimination, and
the double-$\cPqb$ tag discriminant and $m_{\mathrm{SD}}^\PH$ used for tagging candidate Higgs boson jets.  The QCD distributions are
extracted from events with the generator-level $\HT>1\TeV$.
Each variable distribution in this set of figures requires an event that passes the selection on all other variables in order to preserve possible correlations.
}
\label{figs:NM1plots}

\end{figure}

Figure~\ref{figs:NM1plots} shows the variable distributions that are used for top and Higgs candidate jet tagging in $\ttbar$, QCD, and signal MC simulation.
The selections for these distributions includes all other top and Higgs candidate jet selections in order to preserve variable correlations.

In the rare occurrence that a jet passes both the Higgs and top jet tags, the ambiguity is
resolved by giving the Higgs jet tag priority.

\subsection{\texorpdfstring{$\cPqb$}{b} jet tagging}
The $\cPqb$ quark from the VLQ or $\PWpr$ decay is reconstructed as an AK4 jet that is required to pass
the CSVv2 $\cPqb$ tagging algorithm~\cite{Sirunyan:2017ezt} at the same operating point as is used for the subjets of the merged top jet.
A MC to data scale factor for the $\cPqb$ tagging requirement is used in order to
improve the agreement of data and simulation.

\subsection{Event selection}
Event selection details can be found in Table~\ref{table:regions}.  The signal region used for setting cross section upper limits
is required to contain a top, a Higgs boson, and a $\cPqb$ tagged jet.

The sensitivity of the selections used in the analysis have been studied both in the context of the expected
limit and the $\PWpr$ discovery potential.  After identifying the top, Higgs, and $\cPqb$ candidate jets,
the $\PWpr$ candidate mass is analyzed as the invariant mass of the three jets.  Table~\ref{table:efft} shows
the signal efficiency for all samples considered in the analysis.

\begin{table}[ht]
\topcaption{Selection regions used in the analysis.  Tagging discriminator selections and regions described in the text are explicitly defined here.
The signal region (SR) is used to set cross section upper limits, the control regions (CRN)
are used to estimate the QCD background, and the validation region (VR) is used to validate the background estimation procedure.
}
\centering
\begin{tabular}{ll}
Label & Discriminator selections\\
\hline
$\PH_{\text{tag}}$ & $\text{Dbtag}>0.8$ and $105<m_{\mathrm{SD}}^\PH<135\GeV$\\
$\cPqt_{\text{tag}}$ & $SJ_{\text{csvmax}}>0.5426$ and $\tau_3/\tau_2<0.8$ and $105<m_{\mathrm{SD}}^\cPqt<210\GeV$\\
$\cPqb_{\text{tag}}$ & $\text{CSVv2}>0.5426$\\
$\PH_{\text{antitag}}$ & $m_{\mathrm{SD}}^\PH<30\GeV$\\
$\cPqt_{\text{antitag}}$ & $SJ_{\text{csvmax}}>0.5426$ and $\tau_3/\tau_2>0.65$ and $30<m_{\mathrm{SD}}^\cPqt<105\GeV$\\
$\cPqb_{\text{antitag}}$ & $\text{CSVv2}<0.5426$\\
\end{tabular}
\\[3ex]
\begin{tabular}{clll}
\multicolumn{4}{c}{Signal region} \\
\hline
Region & Top jet & Higgs jet & $\cPqb$ jet\\
\hline
SR & $\cPqt_{\text{tag}}$ & $\PH_{\text{tag}}$   &  $\cPqb_{\text{tag}}$\\[\cmsTabSpace]
\multicolumn{4}{c}{Background estimation} \\
\hline
Region & Top jet & Higgs jet & $\cPqb$ jet\\
\hline
CR1 & $\cPqt_{\text{antitag}}$ & $\PH_{\text{antitag}}$   &  $\cPqb_{\text{tag}}$  \\
CR2 & $\cPqt_{\text{antitag}}$ & $\PH_{\text{tag}}$   &  $\cPqb_{\text{tag}}$ \\
CR3 & $\cPqt_{\text{tag}}$ & $\PH_{\text{antitag}}$   &  $\cPqb_{\text{tag}}$  \\
\end{tabular}
\hspace{0.05\textwidth}
\begin{tabular}{clll}
\multicolumn{4}{c}{Validation region} \\
\hline
Region & Top jet & Higgs jet & $\cPqb$ jet\\
\hline
VR & $\cPqt_{\text{tag}}$ & $\PH_{\text{tag}}$   &  $\cPqb_{\text{antitag}}$  \\[\cmsTabSpace]
\multicolumn{4}{c}{Validation background estimation} \\
\hline
Region & Top jet & Higgs jet & $\cPqb$ jet\\
\hline
CR4 & $\cPqt_{\text{antitag}}$ & $\PH_{\text{antitag}}$   &  $\cPqb_{\text{antitag}}$   \\
CR5 & $\cPqt_{\text{antitag}}$ & $\PH_{\text{tag}}$   &  $\cPqb_{\text{antitag}}$ \\
CR6 & $\cPqt_{\text{tag}}$ & $\PH_{\text{antitag}}$   &  $\cPqb_{\text{antitag}}$  \\
\end{tabular}
\label{table:regions}
\end{table}

\begin{table}[ht]
\topcaption{The selection efficiency (\%) for each signal mass point in the analysis.}
\label{table:efft}
\centering
\begin{tabular}{rcccccc}
&\multicolumn{6}{c}{$m_\mathrm{\PWpr}(\GeVns{})$}\\ \cline{2-7}
   $m_\mathrm{VLQ}(\GeVns{})$  & 1500 & 2000 & 2500 & 3000 & 3500 & 4000\\
\hline
 800 & 0.70 $\pm$ 0.13 &  &  &   &  &   \\
 1000 & 0.91 $\pm$ 0.18 & 2.3 $\pm$ 0.4 &  &   &  &   \\
 1300 & 0.48 $\pm$ 0.09 & 2.6 $\pm$ 0.4 & 3.7 $\pm$ 0.6 &   &  &   \\
 1500 &  & 2.1 $\pm$ 0.4 & 3.7 $\pm$ 0.6 & 4.2 $\pm$ 0.7 &  &  \\
 1800 &  &  & 3.2 $\pm$ 0.5 & 4.1 $\pm$ 0.7  & 4.4 $\pm$ 0.7 &   \\
 2100 &  &  &  & 3.7 $\pm$ 0.6  & 4.2 $\pm$ 0.7 & 4.4 $\pm$ 0.7 \\
 2500 &  &  &  &  & 3.8 $\pm$ 0.6 & 4.0 $\pm$ 0.7 \\
 3000 &  &  &  &  &  &  3.4 $\pm$ 0.6 \\
\end{tabular}
\end{table}

\section{Background estimation}
\label{sec:backgroundestimation}

The primary background in this analysis is QCD multijet production, the contribution of which is derived from data using control regions that are selected
with kinematic criteria that are similar to those used for the signal region but with a reduced signal efficiency.
This is achieved by
inverting top substructure selections and extracting the Higgs jet pass to fail ratio for QCD jets.  This ratio is then used as an event weight
for events that pass the top jet selection but fail the Higgs boson jet selection.  The resulting distribution is used as the background estimate for the signal region.
The primary assumption for the background estimate method is that the top jet substructure selection can be inverted without
largely biasing the Higgs jet substructure selection.

\begin{figure}[htb]
\centering
\includegraphics[width=0.48\textwidth]{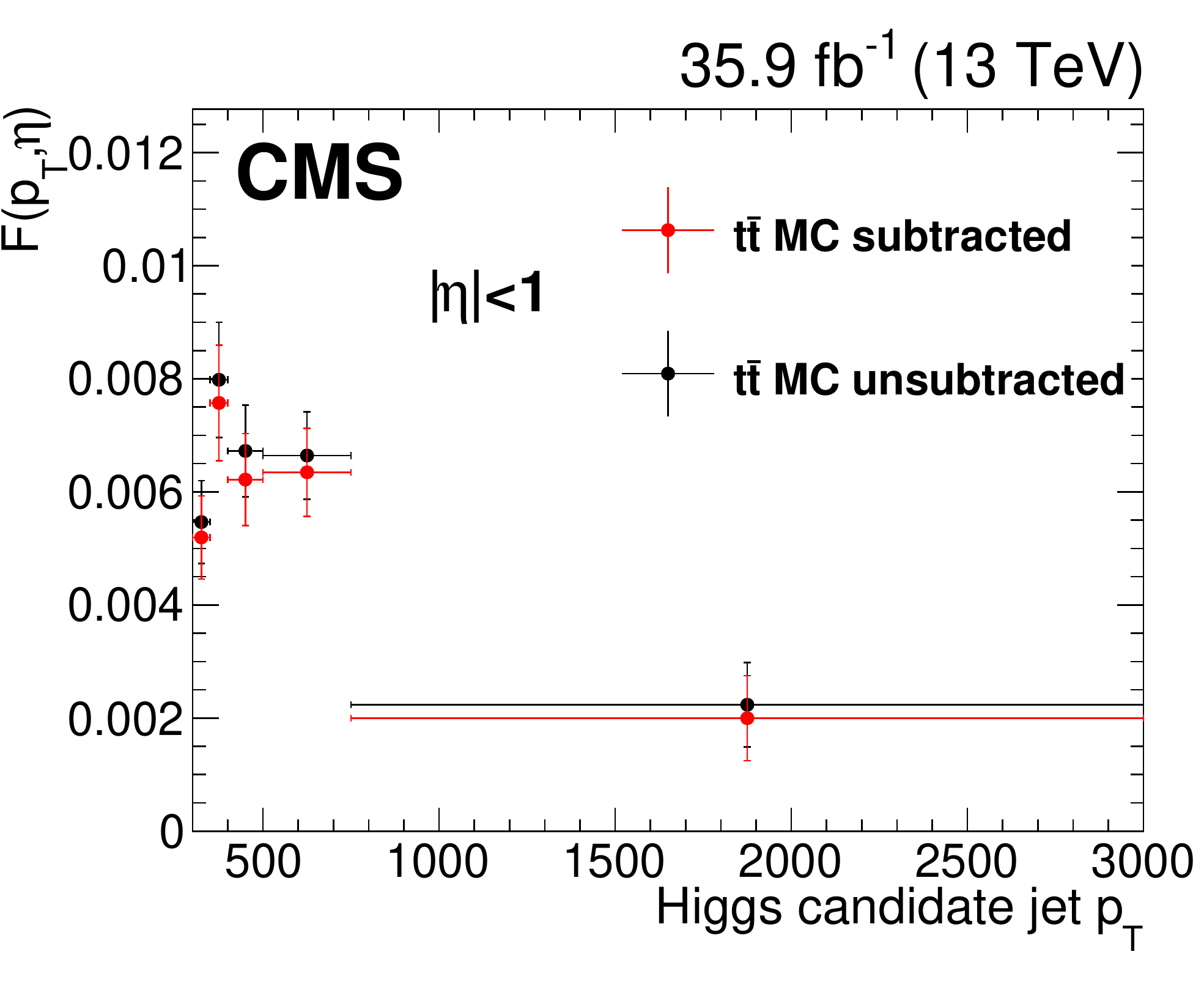}
\includegraphics[width=0.48\textwidth]{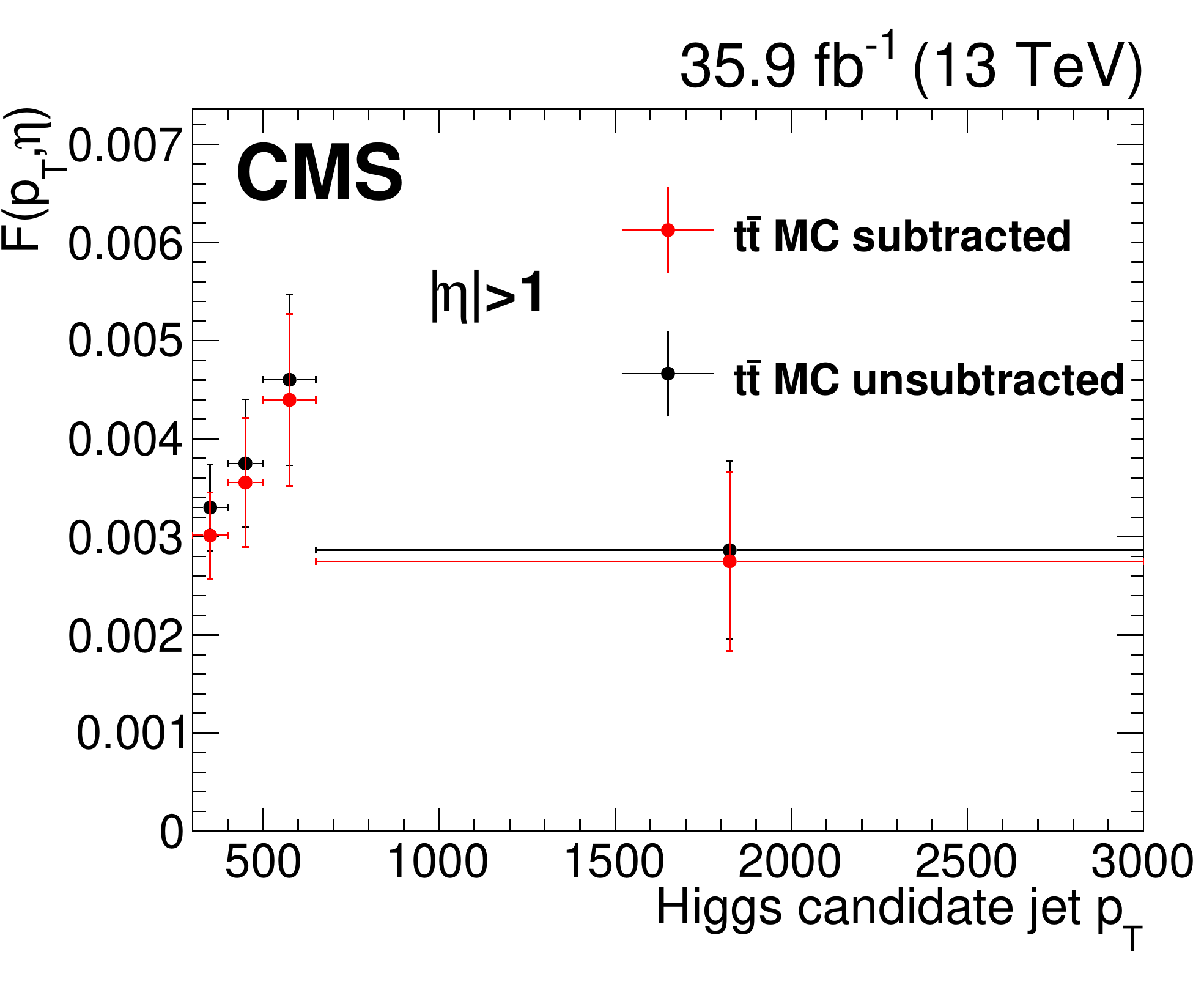}
\caption{
Transfer function $\TF$ used for estimation of the QCD background in the signal region, shown in the central (left) and forward (right) $\eta$ regions.
The error bars represent the statistical uncertainty in $\TF$ only.
}
\label{figs:SRhrate}

\end{figure}

A set of control regions are defined by requiring the Higgs jet candidate $m_{\mathrm{SD}}^\PH$ to be less than 30$\GeV$ with
no double-$\cPqb$ tagging selection.
Table~\ref{table:regions} defines various selection regions used in the analysis.  A transfer function $\TF$ is extracted from
data by inverting the top jet candidate $m_{\mathrm{SD}}^\cPqt$ selection to be between 30 and 105$\GeV$ and
$\tau_3$/$\tau_2>0.65$.
In this region, $\TF$ is defined as the ratio of the jet $\pt$ spectrum of the tagged Higgs candidate in two $\eta$ regions
(central, $\abs{\eta}<1.0$, and forward, $\abs{\eta}>1.0$) for the full Higgs jet selection (CR2) to the
inverted selection (CR1) and is shown in Fig.~\ref{figs:SRhrate}.  The $\TF$ distribution is used to transform the
normalization and shape of distributions from the $\PH_{\text{antitag}}$ region to the $\PH_{\text{tag}}$ selection region, and
is measured with low signal contamination.

The $\TF$ function is then used to predict the background in the signal region.  This is accomplished by defining a control region in data
with identical top and $\cPqb$ jet candidate selections as in the signal region, but with the inverted Higgs jet selection (CR3).  In this region,
the $\mthb$ template is created using $\TF$ as an event weight in a given Higgs candidate jet $\pt$, $\eta$ bin.
This weighted template is used as the QCD background estimate in the signal region.

In the $\TF$ extraction procedure, the $\ttbar$ production component is subtracted from data in all distributions used for
creating $\TF$ in order to ensure that $\TF$ refers only to the QCD component.  The fraction of $\ttbar$ simulation subtracted from
the numerator and denominator is low, 7.3 and 0.4\% of the total distribution, respectively. Additionally, the $\ttbar$ contamination
of the QCD background estimate in the signal region must to be subtracted.  This is performed by applying the QCD background estimation procedure to simulated
$\ttbar$ events using the same $\TF$ as is used when extracting the QCD estimate from data.
The estimated contribution accounts for 2.6\% of the total QCD estimate in the signal region, which is then subtracted when forming the background estimate.
The $\ttbar$ contamination has only a small effect on the QCD background estimation, so
the systematic uncertainty due to the $\ttbar$ subtraction procedure is conservatively taken as the difference between the QCD background estimate
extracted with and without the full $\ttbar$ subtraction procedure.

\begin{figure}[htb]
\centering
\includegraphics[width=0.7\textwidth]{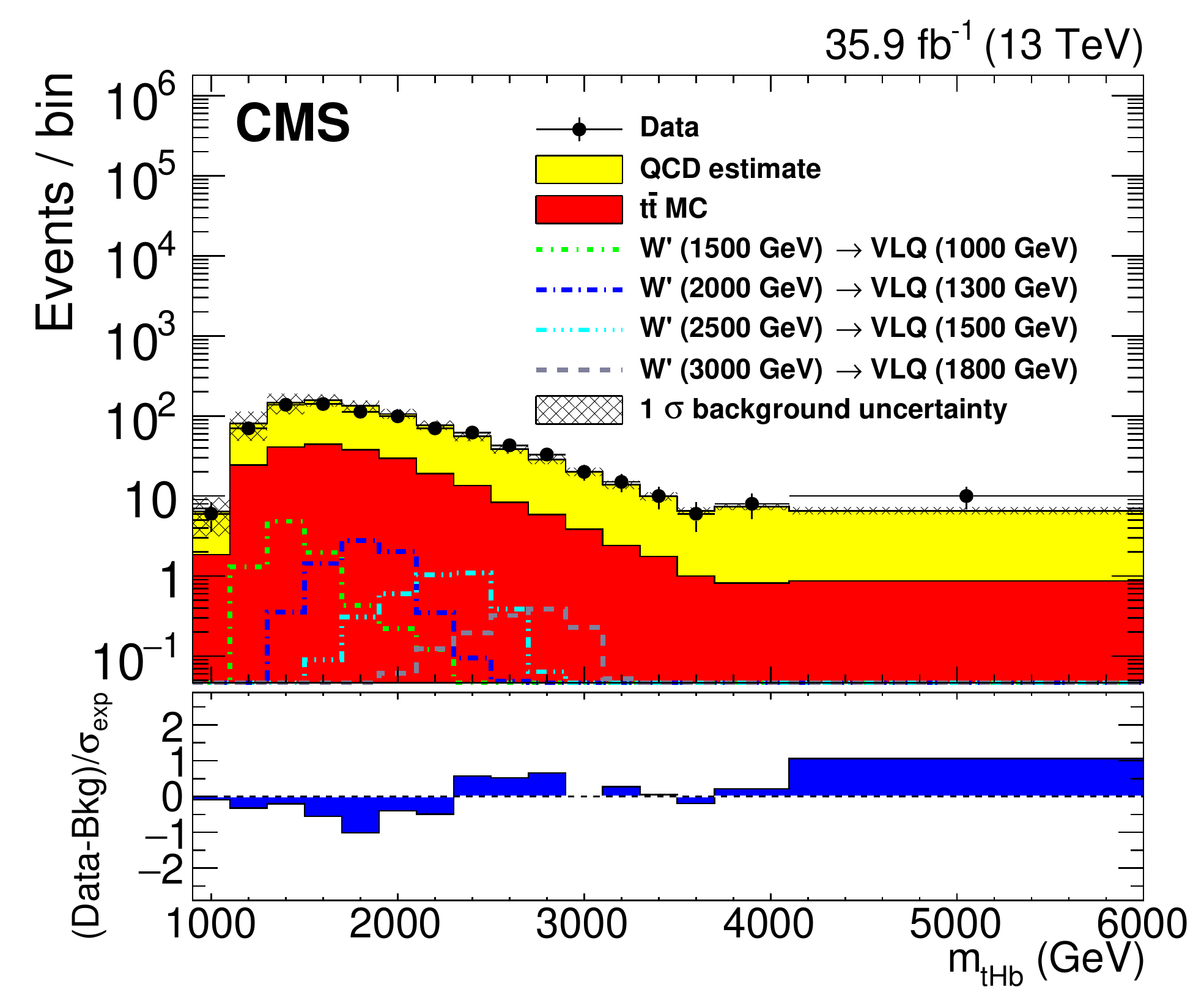}
\caption{
Reconstructed $\PWpr$ mass distributions ($\mthb$) in the $\cPqb$ candidate inverted validation region (VR) shown for data and background contributions.
Several signal hypotheses are shown to demonstrate the low signal contamination.  The background uncertainty
includes all systematic and statistical uncertainties.
}
\label{figs:wpmass}

\end{figure}

In order to test the applicability and versatility of the background estimate in data, a validation region, VR as defined in Table~\ref{table:regions}, is defined
based on inverting the
$\cPqb$ tagging criterion on the $\cPqb$ candidate jet, with the corresponding control regions for background estimation (CR4--CR6).  The transfer function in this validation region $\TFV$
is estimated from the ratio of CR5 to CR4 using the same parameterization as $\TF$.
The $\mthb$ background validation test in this region can be seen in Fig.~\ref{figs:wpmass}.
This region validates the background estimate analog with a $\chi^2$/ndf of 0.3 with systematic uncertainties taken into account, where
ndf is the number of degrees of freedom.
The $\ttbar$ component in this validation region is removed using the same procedure that is used in the signal region background estimate.
The agreement in the $\mthb$ distribution background validation test demonstrates that the top jet selection
can be inverted without biasing the Higgs jet selection.  The Higgs jet candidate 4-vector mass for the SR background estimate
is set to the mean of the distribution extracted from the VR in order to correct the small kinematic bias
from the mass selection when forming the $\mthb$ invariant mass.
This correction has only a small effect on the resulting distribution because of the fact
that the jet $\pt$ is large compared to the mass, and a systematic uncertainty is evaluated as the root mean square of the distribution in the VR.

\begin{figure}[htb]
\centering
\includegraphics[width=0.7\textwidth]{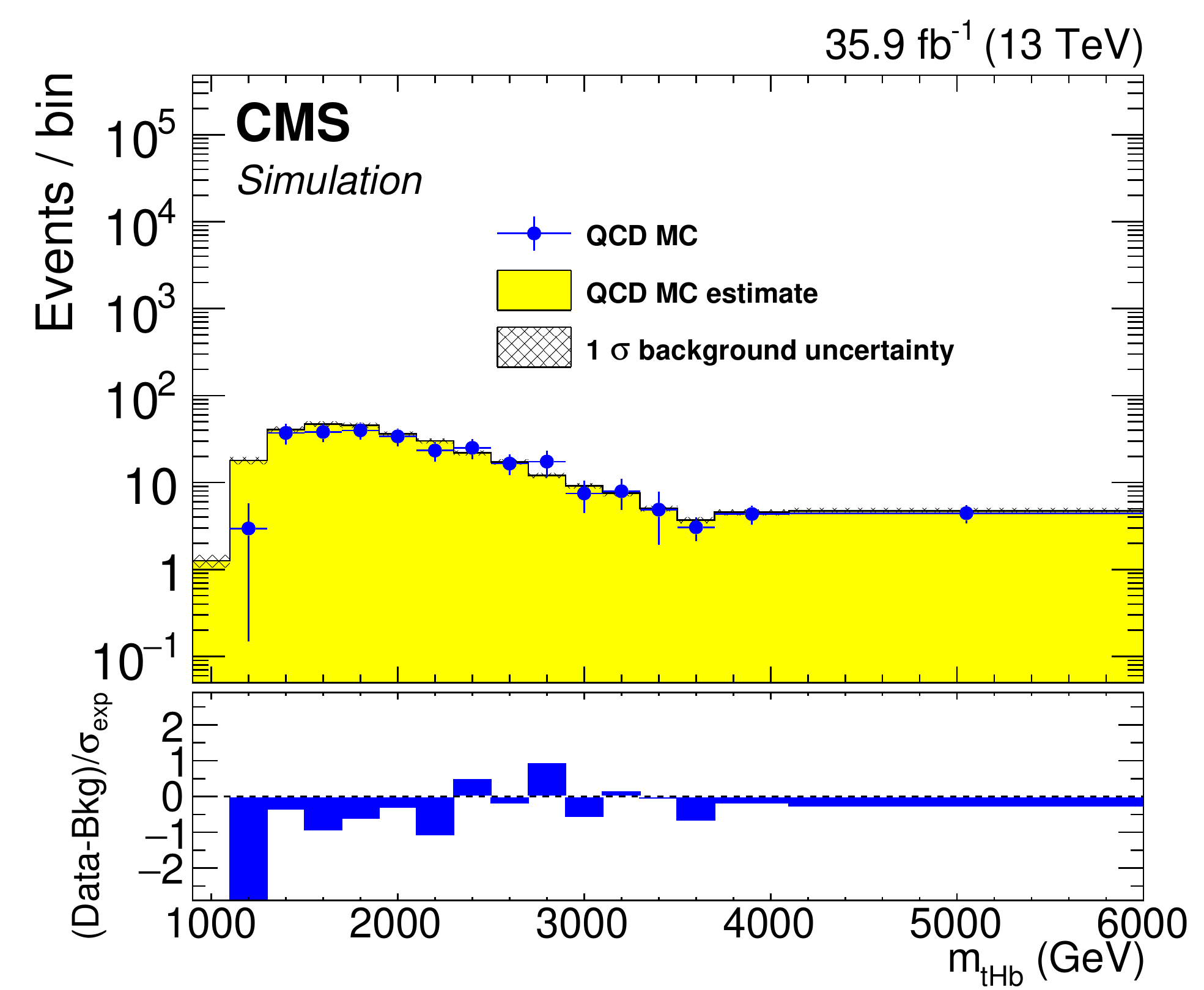}
\caption{
Reconstructed $\PWpr$ mass distributions ($\mthb$) for the simulated QCD events in the signal region for the purposes of validation.  The agreement
given the systematic uncertainties is at the 1 standard deviation level.  The background uncertainty
takes into account all systematic and statistical uncertainties.
}
\label{figs:SRwpmassQCD}
\end{figure}

Additionally, the background validation can be studied with simulated QCD events.  Figure \ref{figs:SRwpmassQCD} shows
 the level of background agreement where the SR selection and QCD background are evaluated using only simulated QCD events with the same method as was previously described for data.
A $\chi^2$/ndf of 1.4 is observed, and
an additional systematic uncertainty is included when evaluating the QCD background
estimate in collision data.  This correction is extracted from the ratio of the SR to QCD background in the QCD MC validation test,
and is applied using an interpolation of the ratio in order to decrease the effect of statistical
fluctuations but to still keep the increased uncertainty at low $\mthb$.

The $\ttbar$ component is estimated by using simulation with an additional event weight to correct the generator
top jet $\pt$ distribution~\cite{PhysRevD.95.092001}.  This generator correction is used in order to improve
the agreement of MC with data with respect to a known generator level mismodelling and is cross checked in the VR region.

\section{Systematic uncertainties}
\label{sec:systematics}
This analysis considers a wide range of systematic uncertainties that are organized into those that impact only the event yields, which
are assumed to follow a log-normal distribution \cite{Conway:2011in},
and those that affect the $\mthb$ distribution shape as well. All of the systematic uncertainties
considered in the analysis are summarized in Table~\ref{table:syst}.

\subsection{Normalization uncertainties}

The uncertainty in the integrated luminosity is taken as 2.5\% for the data set used in the analysis~\cite{CMS-PAS-LUM-17-001}.

The uncertainty in the correction to the efficiency of top jet tagging algorithm is between $-$4 and $+$10\% of the nominal value.

The theoretical uncertainty in the $\ttbar$ production cross section
is taken into account as an asymmetric uncertainty between $-$5.5 and $+$4.8\% that
is calculated as the quadrature sum of the scale and PDF uncertainties on the overall cross section.

\subsection{Shape uncertainties}

The uncertainty in the jet energy scale is taken into account by scaling the four-vectors
used in reconstructing the $\mthb$ distribution by the $\pm$1$\sigma$ jet energy scale uncertainty, which is approximately 2\% for jets in the analysis.  The jet energy scale variation impacts the
$\mthb$ distribution shape through a horizontal shift but can also cause a normalization difference in the case that the
jet falls above or below the kinematic threshold.  The uncertainty in the jet energy resolution is also taken into account
by the $\pm$1$\sigma$ uncertainty in the jet energy resolution correction used for simulated samples.  This uncertainty is applied
to all simulated samples used in the analysis, and has only a small impact.

The uncertainty in the jet mass scale and resolution is measured in a highly enriched sample of $\ttbar$ containing one final state lepton.
In this sample, a fit is performed to the $\PW$ boson jet mass peak in the corresponding AK8 jet PUPPI $m_{\mathrm{SD}}^\PH$ distribution,
in which the mean and width of the PUPPI $m_{\mathrm{SD}}^\PH$ spectrum is extracted.
The mass scale uncertainty is estimated from the shift of the $\PW$ mass peak to be 0.94\%.
The uncertainty in the mass resolution is estimated from the $\PW$ boson mass peak width to be 20\%.
These uncertainties are applied to the signal estimate used in the analysis,
 and result in approximately 4 and 6\% variations in the overall yield for the scale and resolution uncertainties, respectively.
The differences in the $\PW$ and Higgs boson tagging efficiencies are estimated from a comparison of parton showering methods and
are found to be between 4--5\%, so an additional 5\% uncertainty is included for the signal simulated samples used in the analysis.

The uncertainty used for the $\cPqb$ tagging requirement on the AK4 jet is evaluated by varying the $\cPqb$ tagging and $\cPqb$ mistagging scale factor within their
 $\pm$1$\sigma$ uncertainty and are considered uncorrelated with each other.  Given the kinematic selection on the AK4 jet, this uncertainty is evaluated in four
$\pt$ regions from 200--1000$\GeV$.  For jets with a $\pt$ outside of this region, the uncertainty is evaluated as twice the uncertainty at
1000$\GeV$.  This uncertainty is applied to all simulated samples used in the analysis, and results in approximately a 2\% effect.

The double-$\cPqb$ tagging uncertainty used for the Higgs jet tagging~\cite{Sirunyan:2017ezt} selection is evaluated by varying the double-$\cPqb$ tagging scale factor
by the $\pm$1$\sigma$ uncertainty.  The scale factor is parameterized using three regions in \pt.  Similar to the AK4 $\cPqb$ tagging
uncertainty, outside of the kinematic range of the scale factor, the uncertainty is evaluated at twice the maximum range.
The double-$\cPqb$ tagging scale factor uncertainty results in
approximately a 5\% effect.  Also
evaluated is the mistag scale factor in the case of a Higgs boson mistagged as a top quark, as explained in Section~\ref{sec:eventreconstruction}.
The uncertainties in both the Higgs jet tagging efficiency and the mistag rate are applied to all simulated samples used in the analysis, and are treated as
uncorrelated with each other during limit setting.

The events used by the analysis are largely collected where the trigger efficiency is near 100\%, however the small inefficiency
is evaluated using the trigger efficiency extracted from data as parameterized in
$\HT$ (see Fig.~\ref{figs:Trigger}), and the uncertainty is evaluated as half of this inefficiency.
This uncertainty is small ($<$1\%), and is applied to all simulated samples used in the analysis.

As mentioned in Section~\ref{sec:datasample}, the simulated pileup distribution is reweighted to match data using an effective
total inelastic cross section of 69.2$\unit{mb}$.  The uncertainty in this procedure is evaluated by varying the total inelastic cross section
by $\pm$4.6\%~\cite{Sirunyan:2018nqx}.  This uncertainty is applied to all simulated samples used in the analysis, and has only a
small impact.

The $\mthb$ distribution from the $\ttbar$ simulation is reweighted to correct for known differences
in the generator $\pt$ spectrum~\cite{PhysRevD.95.092001}.  The $\pm$1$\sigma$ shape uncertainty in this procedure is estimated from the difference with the
unweighted distribution.  This uncertainty is applied to the $\ttbar$ simulated sample used in the analysis, and results in
approximately a 21\% effect.

The PDF uncertainty is evaluated using the NNPDF3.0 set~\cite{Ball:2017nwa}.
The NNPDF set uses MC replicas, from which the uncertainty is evaluated as the RMS of the
distribution of the associated weights, and is then added in quadrature with the $\alpha_s$ uncertainty.
In the case of signal, the shapes are then normalized to the nominal distribution, as to only preserve the shape
of the PDF uncertainty.  The normalization component of the PDF uncertainty is considered an uncertainty in the
signal cross section.

The renormalization and factorization ($\mu_\mathrm{R}$ and $\mu_\mathrm{F}$) scale uncertainty is evaluated using event weights provided for
varying the $\mu_\mathrm{R}$ or $\mu_\mathrm{F}$ scales up and down by a factor of two.  There are six total weights that represent
the independent and simultaneous variation of $\mu_\mathrm{R}$ and $\mu_\mathrm{F}$.  Per event,
all weights are considered and the envelope is then used as the $\pm$1$\sigma$ uncertainty band.
This uncertainty is applied to the $\ttbar$ MC sample used in the analysis, and results in an approximately 30\% effect.
Similar to the PDF uncertainty,
the normalization component of this uncertainty is taken as the signal cross section theoretical uncertainty, and the shape component alone is used for limit setting.

The analysis considers five sources of uncertainty in the shape of the QCD background estimate derived from data.  The statistical uncertainty in
$\TF$ is propagated to the $\mthb$ spectrum by evaluating the $\TF$ weight at $\pm$1$\sigma$ in a given ($\pt$, $\eta$) bin.
The uncertainty from each $\TF$ bin is added in quadrature to form the full uncertainty in the $\mthb$ template.
The up and down uncertainty variation in the $\ttbar$ subtraction procedure is taken as the unsubtracted $\mthb$ distribution and the
resulting $\mthb$ distribution given twice the subtraction.  The uncertainty in the four vector Higgs jet candidate mass
modification is taken as $\pm$30$\GeV$.  The ``nonclosure'' uncertainty in the QCD background estimate is evaluated as the difference
between the full selection and background prediction from the QCD MC closure test using the interpolated ratio, and is the leading source
of uncertainty in the QCD background estimate of approximately 20\%.

The MC statistical uncertainty is taken into account using the ``Barlow--Beeston lite'' method~\cite{BARLOW1993219} during limit setting.

\begin{table}[htb]
\topcaption{Sources of systematic uncertainty affecting the $\mthb$ distribution.
Sources that list the systematic variation as $\pm 1\sigma$ depend on the distribution of the variable given in the parentheses,
while those that list the variation in percent are rate uncertainties.}
\label{table:syst}
\centering
\begin{tabular}{lcc}
Source & Variation & Process  \\
\hline
Integrated luminosity & $\pm$2.5\% & signal, $\ttbar$ \\
Top jet tagging & $+10.0$\%, $-4$\% &  signal, $\ttbar$ \\
$\ttbar$ cross section & $+4.8$\%, $-5.5$\% & $\ttbar$  \\
Top quark $\pt$ reweighting & $+1\sigma (\pt(\mathrm{gen})) $& $\ttbar$  \\
Matrix element $\mu_\mathrm{R}/\mu_\mathrm{F}$ scales & $\pm 1\sigma (\mu_\mathrm{R}/\mu_\mathrm{F})$ & signal, $\ttbar$  \\
Jet energy scale & $\pm 1\sigma (\pt,\eta)$ & signal, $\ttbar$ \\
Jet energy resolution & $\pm 1\sigma (\pt, \eta)$ & signal, $\ttbar$ \\
Jet mass scale & $\pm 1\sigma (m_{\mathrm{SD}}^\PH)$ & signal, $\ttbar$ \\
Jet mass resolution & $\pm 1\sigma (m_{\mathrm{SD}}^\PH)$ & signal, $\ttbar$ \\
$\cPqb$ tagging & $\pm 1\sigma (\pt)$ &  signal, $\ttbar$ \\
$\cPqb$ mistagging & $\pm 1\sigma (\pt)$ &  signal, $\ttbar$ \\
Double-$\cPqb$ tagging & $\pm 1\sigma (\pt)$ & signal, $\ttbar$  \\
Double-$\cPqb$ mistagging & $\pm 1\sigma (\pt)$ & signal, $\ttbar$  \\
Higgs jet tagging & $\pm$5\% & signal \\
Pileup & $\pm 1\sigma$ ($\sigma_{\mathrm{mb}}$) & signal, $\ttbar$ \\
PDF & $\pm 1\sigma (Q^2,x)$ &  signal, $\ttbar$ \\
$\HT$ trigger & $\pm 1\sigma (\HT)$ & signal, $\ttbar$ \\
$\ttbar$ contamination & $\pm 1\sigma (\pt)$ & QCD \\
$\TF$ & $\pm 1\sigma (\pt, \eta)$ & QCD \\
Higgs jet mass modification & $\pm 1\sigma (m_\mathrm{\PH})$ & QCD  \\
Nonclosure & $\pm 1\sigma (\mthb)$& QCD  \\
\end{tabular}
\end{table}

\section{Results}
\label{sec:limits}

\begin{figure}[htb]
\centering
\includegraphics[width=0.7\textwidth]{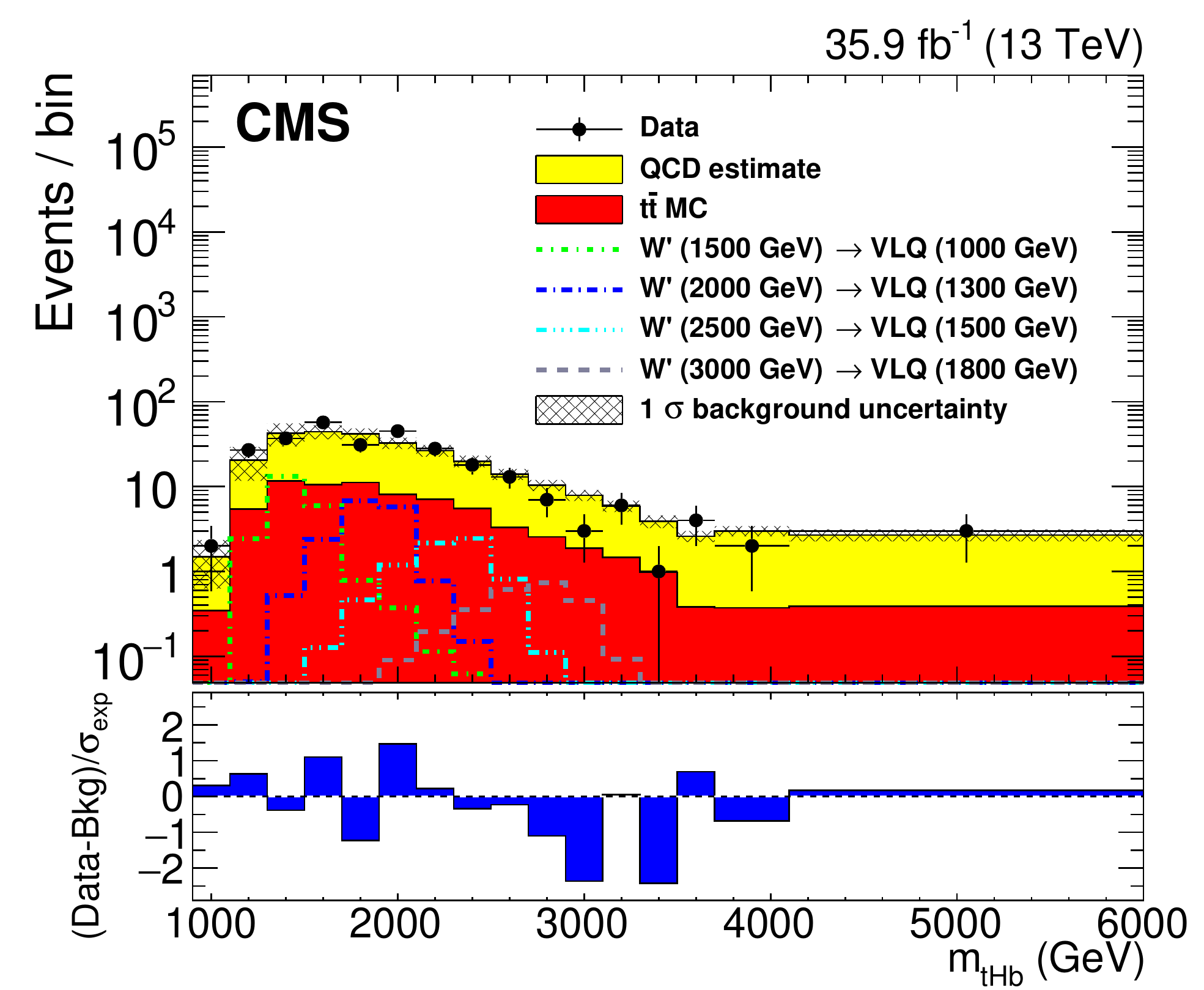}\\
\caption{
Reconstructed $\PWpr$ mass distributions ($\mthb$) in the signal region, compared with the distributions of estimated backgrounds,
and several benchmarks models.  The signal distributions include the contributions from
$\PWpr$ decays to both the $\tpr$ and $\bpr$ assuming equal branching fractions.  The uncertainties
shown in the hatched region contain both statistical and systematic uncertainties of all background components.
}
\label{figs:SRwpmass}
\end{figure}

The final $\mthb$ distribution is shown in Fig.~\ref{figs:SRwpmass}, with a $\chi^2$/ndf of 1.3 for the agreement of data and background.
Table~\ref{table:cutflow} shows the yield for data, QCD and $\ttbar$ backgrounds, for various selection regions including the full selection.

\begin{table}
\centering
\topcaption{Event yield table after various selections.
The definition of each region is given in Table~\ref{table:regions}.  The uncertainties shown here for the
validation region and the signal region are pre fit; the posteriori uncertainties for
$\ttbar$ and QCD are constrained down by 40 and 14\%, respectively. }
\label{table:cutflow}
\begin{tabular}{crrr}
Region & Data & QCD & $\ttbar$  \\
\hline
CR1 & 79\,104 & $\NA$ & 332 \\
CR2 & 398 & $\NA$ & 25 \\
CR3 & 45\,646 & $\NA$ & 1365 \\
CR4 & 288\,926 & $\NA$ & 543 \\
CR5 & 1\,330 & $\NA$ & 76 \\
CR6 & 154\,608 & $\NA$ & 1991 \\
VR & $844\pm 30$ & $659 \pm 150$ & $236 \pm 83$ \\
SR & $284 \pm 17$ & $208 \pm 49$ & $71 \pm 28$ \\
\end{tabular}
\end{table}

Using a Bayesian approach with a flat prior for the signal cross section, upper
limits are obtained on the product of the $\PWpr$ boson production cross section in the $s_{\mathrm{L}}=0.5$ and $\cott=3$ hypothesis,
and the benchmark $\PWpr\to\tpr/\bpr\to\cPqt\PH\cPqb$ branching fraction.
A binned likelihood is used to calculate 95\% confidence level (\CL) upper limits,
in a process where all systematic uncertainties listed in Section 6 that affect the shape
of the $\mthb$ distribution are included as nuisance parameters that modify the shape using template interpolation,
and those that affect the normalization are included as nuisance parameters with lognormal priors.  For the
signal template, the sum of reconstructed $\mthb$ distribution from the \cPqt$\bpr$ and \cPqb$\tpr$ decay channels is used.

\begin{figure}[htb]
\centering
\includegraphics[width=0.48\textwidth]{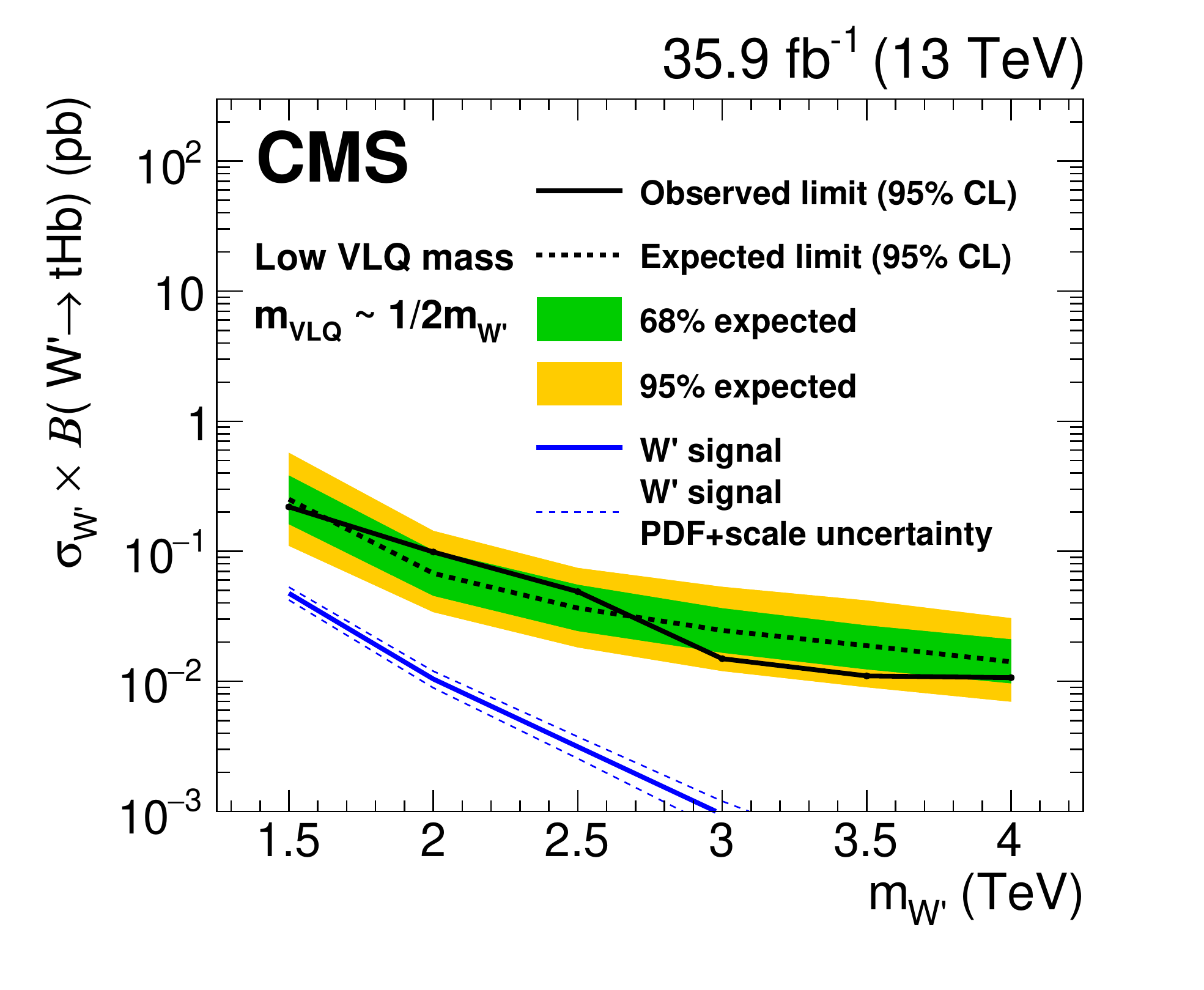}
\includegraphics[width=0.48\textwidth]{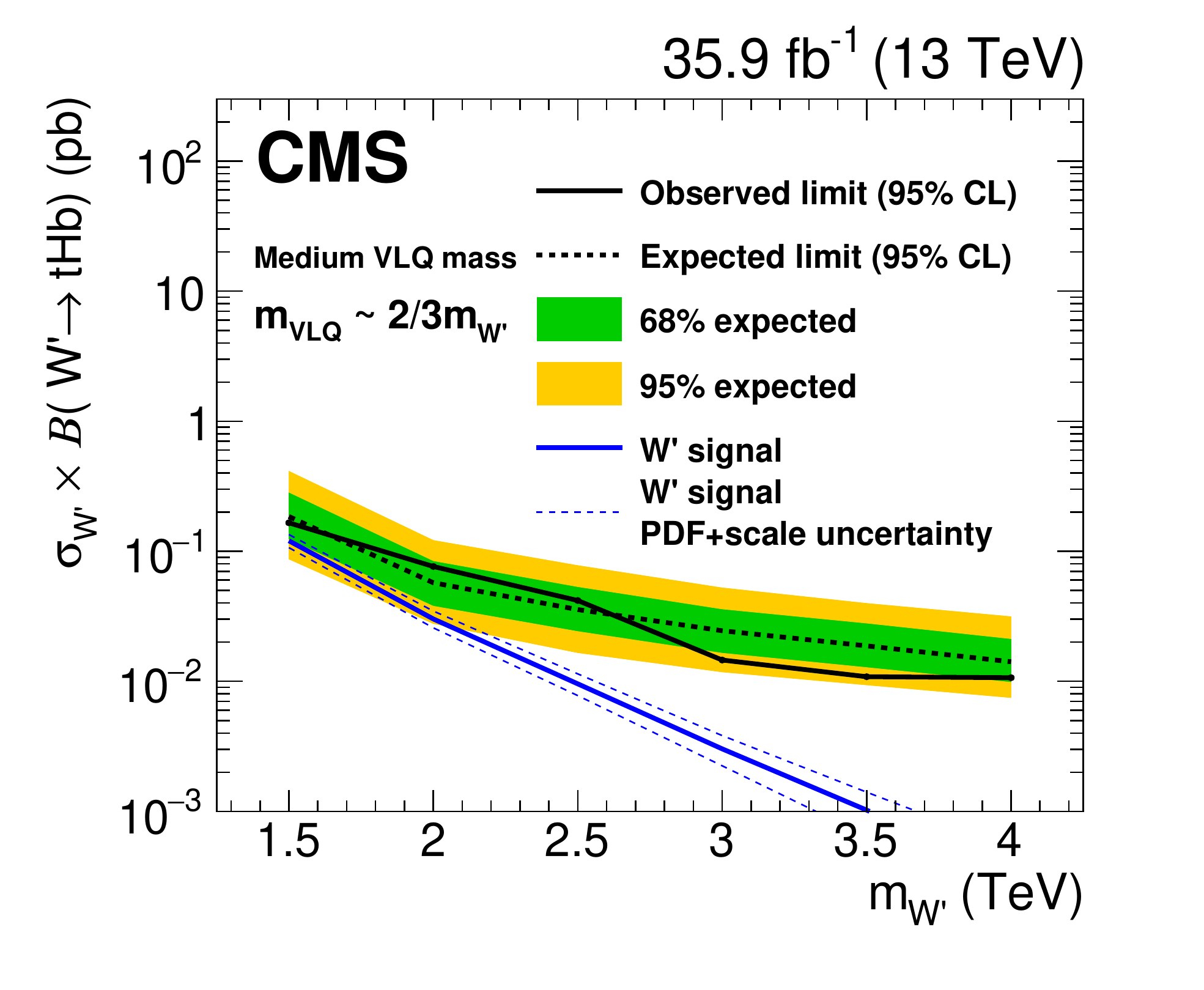}\\
\includegraphics[width=0.48\textwidth]{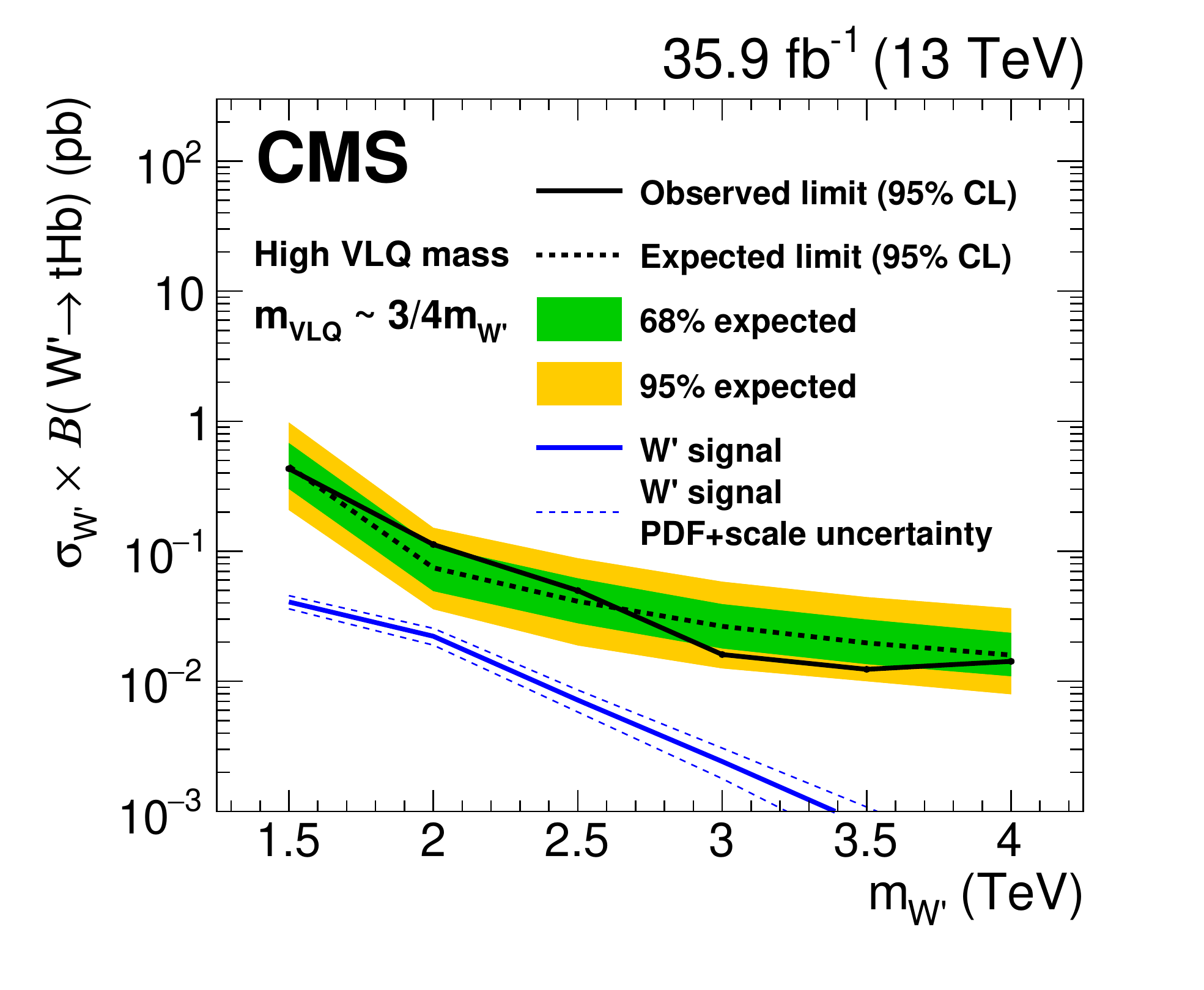}
\caption{
The $\PWpr$ boson 95\% \CL production cross section limits.
The expected limits (dashed) and observed limits (solid), as well as the $\PWpr$ boson theoretical cross section and the PDF and scale
normalization uncertainties are shown.
The bands around the expected limit represent the $\pm$1 and ${\pm}2\sigma_{exp}$ uncertainties in the expected limit.  The limits for low- (upper left),
medium- (upper right), and high- (lower) mass VLQ mass ranges, defined in Table~\ref{table:efft}, are shown.
}
\label{figs:limits}
\end{figure}

Pseudo-experiments are used to derive the ${\pm}1 \sigma$ deviations in the expected limit.
The systematic uncertainties described above are accounted for as nuisance parameters and the posterior probability is refitted for each pseudo-experiment.
Cross section upper limits are shown in Fig.~\ref{figs:limits}.
The highest signal significance is at $M_{\PWpr}=2\TeV$ from the high VLQ mass hypothesis at a value of 0.85 standard deviations.
Although no signal mass points are excluded by solely analyzing the all hadronic $\PWpr\to\tpr/\bpr\to\cPqt\PH\cPqb$ decay in the democratic
$\cPqb\tpr$ and $\cPqt\bpr$ decay hypothesis, a $\PWpr$ with a mass below 1.6\TeV is excluded at 95\% \CL in the case of
a 100\% $\cPqb\tpr$ branching fraction hypothesis.

\section{Summary}
A search for a heavy $\PWpr$ boson decaying to a $\bpr$ or $\tpr$ vector-like quark and a top or $\cPqb$ quark, respectively,
has been presented. The data correspond to an integrated luminosity of 35.9\fbinv collected in 2016 with the CMS detector at the LHC.
The signature considered for both decay modes is a top quark and a Higgs boson, both decaying hadronically, and a $\cPqb$ quark jet.
Boosted heavy-resonance identification techniques are used to exploit the event signature of three energetic jets and to suppress standard model
backgrounds. No significant deviation from the standard model background prediction has been observed.
Cross section upper limits on $\PWpr$ boson production in the top quark, Higgs boson, and $\cPqb$ quark decay mode
are set as a function of the $\PWpr$ mass, for several vector-like quark mass hypotheses.
These are the first limits for $\PWpr$ boson production in this decay channel, and cover a range of 0.01 to 0.43$\unit{pb}$ in the
$\PWpr$ mass range between 1.5 and 4.0\TeV.

\begin{acknowledgments}
We congratulate our colleagues in the CERN accelerator departments for the excellent performance of the LHC and thank the technical and administrative staffs at CERN and at other CMS institutes for their contributions to the success of the CMS effort. In addition, we gratefully acknowledge the computing centers and personnel of the Worldwide LHC Computing Grid for delivering so effectively the computing infrastructure essential to our analyses. Finally, we acknowledge the enduring support for the construction and operation of the LHC and the CMS detector provided by the following funding agencies: BMBWF and FWF (Austria); FNRS and FWO (Belgium); CNPq, CAPES, FAPERJ, FAPERGS, and FAPESP (Brazil); MES (Bulgaria); CERN; CAS, MoST, and NSFC (China); COLCIENCIAS (Colombia); MSES and CSF (Croatia); RPF (Cyprus); SENESCYT (Ecuador); MoER, ERC IUT, and ERDF (Estonia); Academy of Finland, MEC, and HIP (Finland); CEA and CNRS/IN2P3 (France); BMBF, DFG, and HGF (Germany); GSRT (Greece); NKFIA (Hungary); DAE and DST (India); IPM (Iran); SFI (Ireland); INFN (Italy); MSIP and NRF (Republic of Korea); MES (Latvia); LAS (Lithuania); MOE and UM (Malaysia); BUAP, CINVESTAV, CONACYT, LNS, SEP, and UASLP-FAI (Mexico); MOS (Montenegro); MBIE (New Zealand); PAEC (Pakistan); MSHE and NSC (Poland); FCT (Portugal); JINR (Dubna); MON, RosAtom, RAS, RFBR, and NRC KI (Russia); MESTD (Serbia); SEIDI, CPAN, PCTI, and FEDER (Spain); MOSTR (Sri Lanka); Swiss Funding Agencies (Switzerland); MST (Taipei); ThEPCenter, IPST, STAR, and NSTDA (Thailand); TUBITAK and TAEK (Turkey); NASU and SFFR (Ukraine); STFC (United Kingdom); DOE and NSF (USA).

\hyphenation{Rachada-pisek} Individuals have received support from the Marie-Curie program and the European Research Council and Horizon 2020 Grant, contract No. 675440 (European Union); the Leventis Foundation; the A. P. Sloan Foundation; the Alexander von Humboldt Foundation; the Belgian Federal Science Policy Office; the Fonds pour la Formation \`a la Recherche dans l'Industrie et dans l'Agriculture (FRIA-Belgium); the Agentschap voor Innovatie door Wetenschap en Technologie (IWT-Belgium); the F.R.S.-FNRS and FWO (Belgium) under the ``Excellence of Science - EOS" - be.h project n. 30820817; the Ministry of Education, Youth and Sports (MEYS) of the Czech Republic; the Lend\"ulet (``Momentum") Programme and the J\'anos Bolyai Research Scholarship of the Hungarian Academy of Sciences, the New National Excellence Program \'UNKP, the NKFIA research grants 123842, 123959, 124845, 124850 and 125105 (Hungary); the Council of Science and Industrial Research, India; the HOMING PLUS program of the Foundation for Polish Science, cofinanced from European Union, Regional Development Fund, the Mobility Plus program of the Ministry of Science and Higher Education, the National Science Center (Poland), contracts Harmonia 2014/14/M/ST2/00428, Opus 2014/13/B/ST2/02543, 2014/15/B/ST2/03998, and 2015/19/B/ST2/02861, Sonata-bis 2012/07/E/ST2/01406; the National Priorities Research Program by Qatar National Research Fund; the Programa Estatal de Fomento de la Investigaci{\'o}n Cient{\'i}fica y T{\'e}cnica de Excelencia Mar\'{\i}a de Maeztu, grant MDM-2015-0509 and the Programa Severo Ochoa del Principado de Asturias; the Thalis and Aristeia program cofinanced by EU-ESF and the Greek NSRF; the Rachadapisek Sompot Fund for Postdoctoral Fellowship, Chulalongkorn University and the Chulalongkorn Academic into Its 2nd Century Project Advancement Project (Thailand); the Welch Foundation, contract C-1845; and the Weston Havens Foundation (USA).
\end{acknowledgments}

\clearpage

\bibliography{auto_generated}

\providecommand{\href}[2]{#2}\begingroup\raggedright\begin{thebibliography}{10}%
\makeatletter
\providecommand{\hrefCMSnoop }[0]{\@secondoftwo}%
\makeatother
\providecommand{\doi}{\texttt{doi:}\begingroup \urlstyle{tt}\Url}

\bibitem{doi:10.1146/annurev.nucl.55.090704.151502}
\hrefCMSnoop {}{M.~Schmaltz and D.~Tucker-Smith, ``Little {Higgs} theories'',}
  \textit{ Ann. Rev. of Nucl. and Part. Sci.} \textbf{ 55} (2005) 229,
  \href{http://dx.doi.org/10.1146/annurev.nucl.55.090704.151502}{\doi{10.1146/annurev.nucl.55.090704.151502}},
\href{http://www.arXiv.org/abs/hep-ph/0502182}{\texttt{arXiv:hep-ph/0502182}}.

\bibitem{PhysRevD.64.035002}
\hrefCMSnoop {}{T.~Appelquist, H.-C. Cheng, and B.~A. Dobrescu, ``Bounds on
  universal extra dimensions'',} \textit{ Phys. Rev. D} \textbf{ 64} (2001)
  035002,
  \href{http://dx.doi.org/10.1103/PhysRevD.64.035002}{\doi{10.1103/PhysRevD.64.035002}},
\href{http://www.arXiv.org/abs/hep-ph/0012100}{\texttt{arXiv:hep-ph/0012100}}.

\bibitem{PhysRevD.11.566}
\hrefCMSnoop {}{R.~N. Mohapatra and J.~C. Pati, ``Left-right gauge symmetry and
  an 'isoconjugate' model of {CP} violation'',} \textit{ Phys. Rev. D} \textbf{
  11} (1975) 566,
  \href{http://dx.doi.org/10.1103/PhysRevD.11.566}{\doi{10.1103/PhysRevD.11.566}}.

\bibitem{wpCMS1}
\hrefCMSnoop {}{{CMS Collaboration}, ``Search for heavy gauge {W'} boson in
  events with an energetic lepton and large missing transverse momentum at $
  \sqrt{s} = $ 13 {TeV}'',} \textit{ Phys. Lett. B} \textbf{ 770} (2017)
  \href{http://dx.doi.org/10.1016/j.physletb.2017.04.043}{\doi{10.1016/j.physletb.2017.04.043}},
\href{http://www.arXiv.org/abs/1612.09274}{\texttt{arXiv:1612.09274}}.

\bibitem{Aaboud:2017efa}
\hrefCMSnoop {}{{ATLAS Collaboration}, ``{Search for a new heavy gauge boson
  resonance decaying into a lepton and missing transverse momentum in 36
  fb$^{-1}$ of $pp$ collisions at $\sqrt{s} =$ 13 TeV with the ATLAS
  experiment}'',} \textit{ Eur. Phys. J. C} \textbf{ 78} (2018) 401,
  \href{http://dx.doi.org/10.1140/epjc/s10052-018-5877-y}{\doi{10.1140/epjc/s10052-018-5877-y}},
\href{http://www.arXiv.org/abs/1706.04786}{\texttt{arXiv:1706.04786}}.

\bibitem{Sirunyan:2018iff}
\hrefCMSnoop {}{{CMS Collaboration}, ``{Search for a heavy resonance decaying
  to a pair of vector bosons in the lepton plus merged jet final state at $
  \sqrt{s}=13 $ TeV}'',} \textit{ JHEP} \textbf{ 05} (2018) 088,
  \href{http://dx.doi.org/10.1007/JHEP05(2018)088}{\doi{10.1007/JHEP05(2018)088}},
\href{http://www.arXiv.org/abs/1802.09407}{\texttt{arXiv:1802.09407}}.

\bibitem{wpAtlas1}
\hrefCMSnoop {}{{ATLAS Collaboration}, ``Search for {WW/WZ} resonance
  production in $\nu$qq final states in pp collisions at {$\sqrt{s}$}=13 {TeV}
  with the {ATLAS} detector'',} \textit{ JHEP} \textbf{ 03} (2018) 042,
  \href{http://dx.doi.org/10.1007/JHEP03(2018)042}{\doi{10.1007/JHEP03(2018)042}},
  \href{http://www.arXiv.org/abs/1710.07235}{\texttt{arXiv:1710.07235}}.

\bibitem{wpCMS3}
\hrefCMSnoop {}{{CMS Collaboration}, ``{Searches for $\PWpr$ bosons decaying to
  a top quark and a bottom quark in proton-proton collisions at 13 TeV}'',}
  \textit{ JHEP} \textbf{ 08} (2017) 029,
  \href{http://dx.doi.org/10.1007/JHEP08(2017)029}{\doi{10.1007/JHEP08(2017)029}},
\href{http://www.arXiv.org/abs/1706.04260}{\texttt{arXiv:1706.04260}}.

\bibitem{wpAtlas2}
\hrefCMSnoop {}{{ATLAS Collaboration}, ``{Search for $W' \rightarrow tb$ decays
  in the hadronic final state using pp collisions at $\sqrt{s}=13$ TeV with the
  ATLAS detector}'',} \textit{ Phys. Lett. B} \textbf{ 781} (2018) 327,
  \href{http://dx.doi.org/10.1016/j.physletb.2018.03.036}{\doi{10.1016/j.physletb.2018.03.036}},
\href{http://www.arXiv.org/abs/1801.07893}{\texttt{arXiv:1801.07893}}.

\bibitem{vlqs1}
\hrefCMSnoop {}{{CMS Collaboration}, ``{Search for single production of a
  vector-like T quark decaying to a Z boson and a top quark in proton-proton
  collisions at $\sqrt{s}~=~$13 TeV}'',} \textit{ Phys. Lett. B} \textbf{ 781}
  (2018) 574,
  \href{http://dx.doi.org/10.1016/j.physletb.2018.04.036}{\doi{10.1016/j.physletb.2018.04.036}},
  \href{http://www.arXiv.org/abs/1708.01062}{\texttt{arXiv:1708.01062}}.

\bibitem{vlqs2}
\hrefCMSnoop {}{{CMS Collaboration}, ``{Search for single production of
  vector-like quarks decaying to a b quark and a Higgs boson}'',} \textit{
  JHEP} \textbf{ 06} (2018) 031,
  \href{http://dx.doi.org/10.1007/JHEP06(2018)031}{\doi{10.1007/JHEP06(2018)031}},
\href{http://www.arXiv.org/abs/1802.01486}{\texttt{arXiv:1802.01486}}.

\bibitem{vlqs3}
\hrefCMSnoop {}{{ATLAS Collaboration}, ``{Search for pair- and
  single-production of vector-like quarks in final states with at least one $Z$
  boson decaying into a pair of electrons or muons in $pp$ collision data
  collected with the ATLAS detector at $\sqrt{s} = 13$ TeV}'',} (2018).
\href{http://www.arXiv.org/abs/1806.10555}{\texttt{arXiv:1806.10555}}.

\bibitem{vlqs4}
\hrefCMSnoop {}{{CMS Collaboration}, ``{Search for single production of
  vector-like quarks decaying into a b quark and a W boson in proton-proton
  collisions at $\sqrt s =$ 13 TeV}'',} \textit{ Phys. Lett. B} \textbf{ 772}
  (2017) 634,
  \href{http://dx.doi.org/10.1016/j.physletb.2017.07.022}{\doi{10.1016/j.physletb.2017.07.022}},
\href{http://www.arXiv.org/abs/1701.08328}{\texttt{arXiv:1701.08328}}.

\bibitem{vlqp1}
\hrefCMSnoop {}{{CMS Collaboration}, ``{Search for pair production of
  vector-like quarks in the bWbW channel from proton-proton collisions at
  $\sqrt{s}~=~$13 TeV}'',} \textit{ Phys. Lett. B} \textbf{ 779} (2018) 82,
  \href{http://dx.doi.org/10.1016/j.physletb.2018.01.077}{\doi{10.1016/j.physletb.2018.01.077}},
  \href{http://www.arXiv.org/abs/1710.01539}{\texttt{arXiv:1710.01539}}.

\bibitem{vlqp2}
\hrefCMSnoop {}{{CMS Collaboration}, ``{Search for vector-like T and B quark
  pairs in final states with leptons at $\sqrt{s} =$ 13 TeV}'',} (2018).
\href{http://www.arXiv.org/abs/1805.04758}{\texttt{arXiv:1805.04758}}.

\bibitem{vlqp3}
\hrefCMSnoop {}{{ATLAS Collaboration}, ``{Combination of the searches for
  pair-produced vector-like partners of the third-generation quarks at
  $\sqrt{s} =$ 13 TeV with the ATLAS detector}'',} (2018).
\href{http://www.arXiv.org/abs/1808.02343}{\texttt{arXiv:1808.02343}}.

\bibitem{AGASHE2005165}
\hrefCMSnoop {}{K.~Agashe, R.~Contino, and A.~Pomarol, ``{The minimal composite
  Higgs model}'',} \textit{ Nucl. Phys. B} \textbf{ 719} (2005) 165,
  \href{http://dx.doi.org/10.1016/j.nuclphysb.2005.04.035}{\doi{10.1016/j.nuclphysb.2005.04.035}},
\href{http://www.arXiv.org/abs/hep-ph/0412089}{\texttt{arXiv:hep-ph/0412089}}.

\bibitem{Barducci2013}
D.~Barducci\hrefCMSnoop {}{ {et~al.}, ``Exploring drell-yan signals from the 4d
  composite higgs model at the lhc'',} \textit{ JHEP} \textbf{ 04} (2013) 152,
  \href{http://dx.doi.org/10.1007/JHEP04(2013)152}{\doi{10.1007/JHEP04(2013)152}},
  \href{http://www.arXiv.org/abs/1210.2927}{\texttt{arXiv:1210.2927}}.

\bibitem{Barducci2016}
\hrefCMSnoop {}{D.~Barducci and C.~Delaunay, ``Bounding wide composite vector
  resonances at the lhc'',} \textit{ JHEP} \textbf{ 02} (2016) 55,
  \href{http://dx.doi.org/10.1007/JHEP02(2016)055}{\doi{10.1007/JHEP02(2016)055}},
  \href{http://www.arXiv.org/abs/1511.01101}{\texttt{arXiv:1511.01101}}.

\bibitem{Vignaroli:2014bpa}
\hrefCMSnoop {}{N.~Vignaroli, ``New {W$'$} signals at the {LHC}'',} \textit{
  Phys. Rev. D} \textbf{ 89} (2014) 095027,
  \href{http://dx.doi.org/10.1103/PhysRevD.89.095027}{\doi{10.1103/PhysRevD.89.095027}},
\href{http://www.arXiv.org/abs/1404.5558}{\texttt{arXiv:1404.5558}}.

\bibitem{CMS-PAS-LUM-17-001}
\href {http://cds.cern.ch/record/2257069}{{CMS Collaboration}, ``{CMS}
  luminosity measurements for the 2016 data taking period'',} CMS Physics
  Analysis Summary CMS-PAS-LUM-17-001, CERN, Geneva, 2017.

\bibitem{Chatrchyan:2008zzk}
\hrefCMSnoop {}{{CMS Collaboration}, ``The {CMS} experiment at the {CERN}
  {LHC}'',} \textit{ JINST} \textbf{ 3} (2008) S08004,
  \href{http://dx.doi.org/10.1088/1748-0221/3/08/S08004}{\doi{10.1088/1748-0221/3/08/S08004}}.

\bibitem{CMS-PRF-14-001}
\hrefCMSnoop {}{{CMS Collaboration}, ``Particle-flow reconstruction and global
  event description with the {CMS} detector'',} \textit{ JINST} \textbf{ 12}
  (2017) P10003,
  \href{http://dx.doi.org/10.1088/1748-0221/12/10/P10003}{\doi{10.1088/1748-0221/12/10/P10003}},
\href{http://www.arXiv.org/abs/1706.04965}{\texttt{arXiv:1706.04965}}.

\bibitem{Cacciari:2008gp}
\hrefCMSnoop {}{M.~Cacciari, G.~P. Salam, and G.~Soyez, ``The anti-$\kt$ jet
  clustering algorithm'',} \textit{ JHEP} \textbf{ 04} (2008) 063,
  \href{http://dx.doi.org/10.1088/1126-6708/2008/04/063}{\doi{10.1088/1126-6708/2008/04/063}},
  \href{http://www.arXiv.org/abs/0802.1189}{\texttt{arXiv:0802.1189}}.

\bibitem{Cacciari:2011ma}
\hrefCMSnoop {}{M.~Cacciari, G.~P. Salam, and G.~Soyez, ``Fastjet user
  manual'',} \textit{ Eur. Phys. J. C} \textbf{ 72} (2012) 1896,
  \href{http://dx.doi.org/10.1140/epjc/s10052-012-1896-2}{\doi{10.1140/epjc/s10052-012-1896-2}},
\href{http://www.arXiv.org/abs/1111.6097}{\texttt{arXiv:1111.6097}}.

\bibitem{Bertolini2014}
\hrefCMSnoop {}{D.~Bertolini, P.~Harris, M.~Low, and N.~Tran, ``{Pileup per
  particle identification}'',} \textit{ JHEP} \textbf{ 10} (2014) 059,
  \href{http://dx.doi.org/10.1007/JHEP10(2014)059}{\doi{10.1007/JHEP10(2014)059}},
\href{http://www.arXiv.org/abs/1407.6013}{\texttt{arXiv:1407.6013}}.

\bibitem{Khachatryan:2016kdb}
\hrefCMSnoop {}{{CMS Collaboration}, ``Jet energy scale and resolution in the
  {CMS} experiment in pp collisions at 8 {TeV}'',} \textit{ JINST} \textbf{ 12}
  (2017) P02014,
  \href{http://dx.doi.org/10.1088/1748-0221/12/02/P02014}{\doi{10.1088/1748-0221/12/02/P02014}},
\href{http://www.arXiv.org/abs/1607.03663}{\texttt{arXiv:1607.03663}}.

\bibitem{Khachatryan:2016bia}
\hrefCMSnoop {}{{CMS Collaboration}, ``{The CMS trigger system}'',} \textit{
  JINST} \textbf{ 12} (2017) P01020,
  \href{http://dx.doi.org/10.1088/1748-0221/12/01/P01020}{\doi{10.1088/1748-0221/12/01/P01020}},
\href{http://www.arXiv.org/abs/1609.02366}{\texttt{arXiv:1609.02366}}.

\bibitem{Powheg}
\hrefCMSnoop {}{S.~Frixione, P.~Nason, and C.~Oleari, ``Matching {NLO QCD}
  computations with parton shower simulations: the {POWHEG} method'',} \textit{
  JHEP} \textbf{ 11} (2007) 070,
  \href{http://dx.doi.org/10.1088/1126-6708/2007/11/070}{\doi{10.1088/1126-6708/2007/11/070}},
\href{http://www.arXiv.org/abs/0709.2092}{\texttt{arXiv:0709.2092}}.

\bibitem{Alioli:2010xd}
\hrefCMSnoop {}{S.~Alioli, P.~Nason, C.~Oleari, and E.~Re, ``A general
  framework for implementing {NLO} calculations in shower {M}onte {C}arlo
  programs: the {POWHEG BOX}'',} \textit{ JHEP} \textbf{ 06} (2010) 043,
  \href{http://dx.doi.org/10.1007/JHEP06(2010)043}{\doi{10.1007/JHEP06(2010)043}},
\href{http://www.arXiv.org/abs/1002.2581}{\texttt{arXiv:1002.2581}}.

\bibitem{Nason:2004rx}
\hrefCMSnoop {}{P.~Nason, ``A new method for combining {NLO QCD} with shower
  {M}onte {C}arlo algorithms'',} \textit{ JHEP} \textbf{ 11} (2004) 040,
  \href{http://dx.doi.org/10.1088/1126-6708/2004/11/040}{\doi{10.1088/1126-6708/2004/11/040}},
\href{http://www.arXiv.org/abs/hep-ph/0409146}{\texttt{arXiv:hep-ph/0409146}}.

\bibitem{Frixione:2007nw}
\hrefCMSnoop {}{S.~Frixione, P.~Nason, and G.~Ridolfi, ``{A positive-weight
  next-to-leading-order Monte Carlo for heavy flavour hadroproduction}'',}
  \textit{ JHEP} \textbf{ 09} (2007) 126,
  \href{http://dx.doi.org/10.1088/1126-6708/2007/09/126}{\doi{10.1088/1126-6708/2007/09/126}},
\href{http://www.arXiv.org/abs/0707.3088}{\texttt{arXiv:0707.3088}}.

\bibitem{Alwall:2014hca}
J.~Alwall\hrefCMSnoop {}{ {et~al.}, ``{The automated computation of tree-level
  and next-to-leading order differential cross sections, and their matching to
  parton shower simulations}'',} \textit{ JHEP} \textbf{ 07} (2014) 079,
  \href{http://dx.doi.org/10.1007/JHEP07(2014)079}{\doi{10.1007/JHEP07(2014)079}},
\href{http://www.arXiv.org/abs/1405.0301}{\texttt{arXiv:1405.0301}}.

\bibitem{Alwall:2007fs}
\hrefCMSnoop {}{J.~Alwall {et~al.}, ``{Comparative study of various algorithms
  for the merging of parton showers and matrix elements in hadronic
  collisions}'',} \textit{ Eur. Phys. J. C} \textbf{ 53} (2008) 473,
  \href{http://dx.doi.org/10.1140/epjc/s10052-007-0490-5}{\doi{10.1140/epjc/s10052-007-0490-5}},
\href{http://www.arXiv.org/abs/0706.2569}{\texttt{arXiv:0706.2569}}.

\bibitem{Sjostrand:2014zea}
T.~Sj{\"o}strand\hrefCMSnoop {}{ {et~al.}, ``An introduction to {PYTHIA}
  8.2'',} \textit{ Comput. Phys. Commun.} \textbf{ 191} (2015) 159,
  \href{http://dx.doi.org/10.1016/j.cpc.2015.01.024}{\doi{10.1016/j.cpc.2015.01.024}},
\href{http://www.arXiv.org/abs/1410.3012}{\texttt{arXiv:1410.3012}}.

\bibitem{CMS-PAS-TOP-16-021}
\href {http://cds.cern.ch/record/2235192}{{CMS Collaboration}, ``Investigations
  of the impact of the parton shower tuning in {Pythia 8} in the modelling of
  $\mathrm{t\overline{t}}$ at $\sqrt{s}=8$ and 13 {TeV}'',} Technical Report
  CMS-PAS-TOP-16-021, CERN, Geneva, 2016.

\bibitem{Khachatryan:2015pea}
\hrefCMSnoop {}{{CMS Collaboration}, ``Event generator tunes obtained from
  underlying event and multiparton scattering measurements'',} \textit{ Eur.
  Phys. J. C} \textbf{ 76} (2016) 155,
  \href{http://dx.doi.org/10.1140/epjc/s10052-016-3988-x}{\doi{10.1140/epjc/s10052-016-3988-x}},
\href{http://www.arXiv.org/abs/1512.00815}{\texttt{arXiv:1512.00815}}.

\bibitem{g4c}
\hrefCMSnoop {}{{GEANT4} Collaboration, ``{\GEANTfour}---a a simulation
  toolkit'',} \textit{ Nucl. Instrum. Meth.} \textbf{ 506} (2003) 250,
  \href{http://dx.doi.org/10.1016/S0168-9002(03)01368-8}{\doi{10.1016/S0168-9002(03)01368-8}}.

\bibitem{CMS-PAS-JME-16-003}
\href {http://cds.cern.ch/record/2256875}{{CMS Collaboration}, ``Jet algorithms
  performance in 13 {TeV} data'',} CMS Physics Analysis Summary
  CMS-PAS-JME-16-003, CERN, Geneva, 2017.

\bibitem{Thaler:2011gf}
\hrefCMSnoop {}{J.~Thaler and K.~Van~Tilburg, ``Maximizing boosted top
  identification by minimizing {N}-subjettiness'',} \textit{ JHEP} \textbf{ 02}
  (2012) 093,
  \href{http://dx.doi.org/10.1007/JHEP02(2012)093}{\doi{10.1007/JHEP02(2012)093}},
\href{http://www.arXiv.org/abs/1108.2701}{\texttt{arXiv:1108.2701}}.

\bibitem{Dasgupta2013}
\hrefCMSnoop {}{M.~Dasgupta, A.~Fregoso, S.~Marzani, and G.~P. Salam,
  ``{Towards an understanding of jet substructure}'',} \textit{ JHEP} \textbf{
  09} (2013) 029,
  \href{http://dx.doi.org/10.1007/JHEP09(2013)029}{\doi{10.1007/JHEP09(2013)029}},
\href{http://www.arXiv.org/abs/1307.0007}{\texttt{arXiv:1307.0007}}.

\bibitem{Larkoski:2014wba}
\hrefCMSnoop {}{A.~J. Larkoski, S.~Marzani, G.~Soyez, and J.~Thaler, ``{S}oft
  {d}rop'',} \textit{ JHEP} \textbf{ 05} (2014)
  \href{http://dx.doi.org/10.1007/JHEP05(2014)146}{\doi{10.1007/JHEP05(2014)146}},
\href{http://www.arXiv.org/abs/1402.2657}{\texttt{arXiv:1402.2657}}.

\bibitem{Sirunyan:2017ezt}
\hrefCMSnoop {}{{CMS Collaboration}, ``{Identification of heavy-flavour jets
  with the CMS detector in pp collisions at 13 TeV}'',} \textit{ JINST}
  \textbf{ 13} (2018) P05011,
  \href{http://dx.doi.org/10.1088/1748-0221/13/05/P05011}{\doi{10.1088/1748-0221/13/05/P05011}},
\href{http://www.arXiv.org/abs/1712.07158}{\texttt{arXiv:1712.07158}}.

\bibitem{PhysRevD.97.072006}
\hrefCMSnoop {}{{CMS Collaboration}, ``{Search for massive resonances decaying
  into $WW$, $WZ$, $ZZ$, $qW$, and $qZ$ with dijet final states at
  $\sqrt{s}=13\text{ }\text{ }\mathrm{TeV}$}'',} \textit{ Phys. Rev. D}
  \textbf{ 97} (2018) 072006,
  \href{http://dx.doi.org/10.1103/PhysRevD.97.072006}{\doi{10.1103/PhysRevD.97.072006}},
\href{http://www.arXiv.org/abs/1708.05379}{\texttt{arXiv:1708.05379}}.

\bibitem{PhysRevD.95.092001}
\hrefCMSnoop {}{{CMS Collaboration}, ``{Measurement of differential cross
  sections for top quark pair production using the lepton+jets final state in
  proton-proton collisions at 13 TeV}'',} \textit{ Phys. Rev. D} \textbf{ 95}
  (2017) 092001,
  \href{http://dx.doi.org/10.1103/PhysRevD.95.092001}{\doi{10.1103/PhysRevD.95.092001}},
\href{http://www.arXiv.org/abs/1610.04191}{\texttt{arXiv:1610.04191}}.

\bibitem{Conway:2011in}
\hrefCMSnoop {}{J.~S. Conway, ``{Incorporating nuisance parameters in
  likelihoods for multisource spectra}'',} in \textit{ {Proceedings, PHYSTAT
  2011 workshop on statistical issues related to discovery claims in search
  experiments and unfolding, CERN,Geneva, Switzerland 17-20 January 2011}},
  p.~115.
\newblock 2011.
\newblock \href{http://www.arXiv.org/abs/1103.0354}{\texttt{arXiv:1103.0354}}.
\newblock
\href{http://dx.doi.org/10.5170/CERN-2011-006.115}{\doi{10.5170/CERN-2011-006.115}}.

\bibitem{Sirunyan:2018nqx}
\hrefCMSnoop {}{{CMS Collaboration}, ``{Measurement of the inelastic
  proton-proton cross section at $ \sqrt{s}=13 $ TeV}'',} \textit{ JHEP}
  \textbf{ 07} (2018) 161,
  \href{http://dx.doi.org/10.1007/JHEP07(2018)161}{\doi{10.1007/JHEP07(2018)161}},
\href{http://www.arXiv.org/abs/1802.02613}{\texttt{arXiv:1802.02613}}.

\bibitem{Ball:2017nwa}
\hrefCMSnoop {}{{NNPDF} Collaboration, ``Parton distributions from
  high-precision collider data'',} \textit{ Eur. Phys. J. C} \textbf{ 77}
  (2017) 663,
  \href{http://dx.doi.org/10.1140/epjc/s10052-017-5199-5}{\doi{10.1140/epjc/s10052-017-5199-5}},
\href{http://www.arXiv.org/abs/1706.00428}{\texttt{arXiv:1706.00428}}.

\bibitem{BARLOW1993219}
\hrefCMSnoop {}{R.~Barlow and C.~Beeston, ``Fitting using finite {M}onte
  {C}arlo samples'',} \textit{ Comput. Phys. Commun.} \textbf{ 77} (1993) 219,
  \href{http://dx.doi.org/10.1016/0010-4655(93)90005-W}{\doi{10.1016/0010-4655(93)90005-W}}.

\end{thebibliography}\endgroup

\cleardoublepage \appendix\section{The CMS Collaboration \label{app:collab}}\begin{sloppypar}\hyphenpenalty=5000\widowpenalty=500\clubpenalty=5000\vskip\cmsinstskip
\textbf{Yerevan Physics Institute, Yerevan, Armenia}\\*[0pt]
A.M.~Sirunyan, A.~Tumasyan
\vskip\cmsinstskip
\textbf{Institut f\"{u}r Hochenergiephysik, Wien, Austria}\\*[0pt]
W.~Adam, F.~Ambrogi, E.~Asilar, T.~Bergauer, J.~Brandstetter, M.~Dragicevic, J.~Er\"{o}, A.~Escalante~Del~Valle, M.~Flechl, R.~Fr\"{u}hwirth\cmsAuthorMark{1}, V.M.~Ghete, J.~Hrubec, M.~Jeitler\cmsAuthorMark{1}, N.~Krammer, I.~Kr\"{a}tschmer, D.~Liko, T.~Madlener, I.~Mikulec, N.~Rad, H.~Rohringer, J.~Schieck\cmsAuthorMark{1}, R.~Sch\"{o}fbeck, M.~Spanring, D.~Spitzbart, A.~Taurok, W.~Waltenberger, J.~Wittmann, C.-E.~Wulz\cmsAuthorMark{1}, M.~Zarucki
% ...
\vskip\cmsinstskip
% ...
1:  Also at Vienna University of Technology, Vienna, Austria\\
% ...
\end{sloppypar}
\end{document}